\documentclass{vldb}
\usepackage{etex}
\providecommand{\mf}[1]{\ensuremath{\mathcal{#1}}}

\providecommand{\algo}{\mbox{\textit{j}LCM}\xspace}
\providecommand{\demoassoc}{\texttt{demo\_assoc}\xspace}
\providecommand{\prodassocreceipt}{\texttt{prod\_assoc\_t}\xspace}
\providecommand{\prodassocclient}{\texttt{prod\_assoc\_c}\xspace}
\providecommand{\prodassocall}{\texttt{prod\_assoc}\xspace}

\providecommand*\circled[1]{\tikz[baseline=(char.base)]{
\node[shape=circle,draw,inner sep=1pt,fill=gray!20] (char) {#1};}}

\providecommand{\project}{\mbox{\textsc{capa}}\xspace}

\providecommand{\comment}[1]{}
\providecommand{\mf}[1]{\ensuremath{\mathcal{#1}}}

\usepackage[utf8]{inputenc}
\usepackage[T1]{fontenc}
\usepackage{graphicx}
\usepackage{balance}
\usepackage[ruled,vlined,linesnumbered,noend]{algorithm2e}
\usepackage[table]{xcolor}
\usepackage{url}
\usepackage{multirow}
\usepackage{hyperref}
\usepackage[caption=false]{subfig}
\usepackage{tikz,pgf}
\usepackage{ctable} 
\usetikzlibrary{positioning,chains,matrix,shapes,shapes.multipart,arrows,shadows,calc,decorations.markings,fit,mindmap,backgrounds}
\usepackage[normalem]{ulem}

\colorlet{g1aColor}{rgb:yellow,1;white,2}
\colorlet{g1bColor}{rgb:yellow,1;white,2}

\colorlet{g2Color}{rgb:red,1;white,2}
\colorlet{g3Color}{rgb:green,1;white,2}
\colorlet{g5Color}{rgb:gray,1;white,2}
\colorlet{g6Color}{rgb:blue,1;white,2}
\colorlet{g4Color}{rgb:brown,1;white,2}

\usepackage[group-separator={,},group-minimum-digits=4]{siunitx}

\DeclareMathOperator{\sign}{sgn}

\newcommand{\titleA}{Testing Interestingness Measures in Practice:\\ A Large-Scale Analysis of Buying Patterns}

\begin{document}

\title{\titleA}

\numberofauthors{2}
\author{
\alignauthor Martin Kirchgessner, Vincent Leroy, Sihem Amer-Yahia, Shashwat Mishra\\
       \affaddr{Univ. Grenoble Alpes, CNRS, LIG}\\
       \affaddr{Grenoble, France}\\
       \email{firstname.lastname@imag.fr}
\alignauthor Intermarch\'e Alimentaire International\\ STIME\\
\affaddr{17 Allée des Mousquetaires}\\
  \affaddr{91078 Bondoufle cedex, France}\\
  \email{contact@mousquetaires.fr}
}

\maketitle

\begin{abstract}
  Understanding customer buying patterns is of great interest to the
  retail industry and has shown to benefit a wide variety of goals
  ranging from managing stocks to implementing loyalty programs. Association rule mining is a
  common technique for extracting correlations such as {\em people in the
  South of France buy ros\'e wine} or {\em customers who buy pat\'e also
  buy salted butter and sour bread.} Unfortunately, sifting through a
  high number of buying patterns is not useful in practice, because
  of the predominance of popular products in the top rules. As a result, a number
  of ``interestingness'' measures (over 30) have been proposed to rank
  rules. However, there is no agreement on which measures
  are more appropriate for retail data.
  Moreover, since pattern mining algorithms output thousands of association rules for each product, the ability for an analyst to rely on ranking measures to identify the most interesting ones is crucial.
  In this paper, we develop \project\ (Comparative Analysis of PAtterns), a framework
  that provides analysts with the ability to compare the outcome of
  interestingness measures applied to buying patterns in the retail
  industry. We report on how we used \project\ to compare 34
  measures applied to over 1,800 stores of
  Intermarch\'e, one of the largest food retailers in France.
\end{abstract}


\section{Introduction} \label{sec:intro}

Ever since databases have been able to store basket data, many
techniques have been proposed to extract useful insights for analysts.
One of the first, association rule mining~\cite{AggrawalSIGMOD1993},
also remains one of the most intuitive.
Association rules are often used to summarize consumer trends in a
transaction\footnote{In this paper, a transaction is an entry in the dataset containing information about the activity of a customer. It never refers to a database transaction.} set or as input to a classifier~\cite{SuganthanSIGMOD2015}. The problem is the very high
number of rules, typically in the order of millions. That is
exacerbated by the lack of thorough studies of which of the many
interestingness measures for ranking rules~\cite{geng2006ACM}
is most appropriate for which application domain. We
present \project, a framework to compare the outcome of different
interestingness measures applied to association rules generated in the
food retail domain. \project\ relies on a flexible architecture and on
\algo~\cite{jlcm}, our parallel and distributed pattern mining
algorithm that runs on MapReduce. The use of real datasets and a close collaboration with
  experienced domain experts from Intermarch\'e, one of the
largest retailers in France, has led us to selecting the most relevant
measures to rank association rules in the food retail domain.

Our dataset contains $290$ million receipts from \num{1884} stores
in all of France, gathered over one year, 2013. Mining this data results
in a huge number of rules.  For example, using a minimum support of
\num{1000} \algo mines \num{2746418} frequent rules of the form
{\em customer segment} $\rightarrow$ {\em product category}.
Out of these, \num{15063} have a confidence of $50\%$ or higher.
Table~\ref{tab:eyecatcher} shows a ranking of the top-10 rules
according to 3 different interestingness measures proposed in ~\cite{geng2006ACM}.
If we denote rules as $A \rightarrow
B$, {\em confidence} is the probability to observe $B$ given that we
observed $A$, i.e., $P(B|A)$. {\em
  Piatetsky-Shapiro}~\cite{Piatetsky1991KDD} combines how $A$ and $B$
occur together with how they would if they were independent, i.e.,
$P(AB) - P(A)P(B)$.  {\em Pearson's} $\chi^2$, measures how unlikely
observations of $A$ and $B$ are independent.  This very small example already shows that these measures
result in different rule rankings.

\begin{table*}
  \setlength{\tabcolsep}{1pt}
  \centering
  \begin{tabular}{|rl|rl|rl|}
    \hline
    \multicolumn{2}{|c|}{\textbf{by confidence}}                 &  \multicolumn{2}{|c|}{\textbf{by Piatetsky-Shapiro~\cite{Piatetsky1991KDD}}} &  \multicolumn{2}{|c|}{\textbf{by Pearson's $\chi^2$}}      \\\hline
    $\{> 65, F, $ Aube$\}$&$\rightarrow$ {\em Dairy}                   & $\{*, *, Nord\}$&$\rightarrow$ {\em Liquids}      & $\{*, *, Somme\}$&$\rightarrow$ {\em Cut cheese}  \\\hline
    $\{> 65, F, $ Aveyron$\}$&$\rightarrow$ {\em Dairy}                & $\{*, *, Nord\}$&$\rightarrow$ {\em Soft drinks}  & $\{*, F, Somme\}$&$\rightarrow$ {\em Cut cheese}   \\\hline
    $\{> 65, F, $ Val de Marne$\}$& $ \rightarrow$ {\em Dairy}         & $\{*, *, Nord\}$&$\rightarrow$ {\em Beers}        & $\{> 65, *, Morbihan\}$&$\rightarrow$ {\em Fresh milk}   \\\hline
    $\{> 65, F, $ Seine S$^{t}$ Denis$\}$& $ \rightarrow$ {\em Dairy}  & $\{*, *, Nord\}$&$\rightarrow$ {\em Spreads}      & $\{> 65, *, Somme\}$&$\rightarrow$ {\em Cut cheese}   \\\hline
    $\{> 65, F, $ Haute Saone$\}$& $ \rightarrow$ {\em Dairy}          & $\{*, F, Nord\}$&$\rightarrow$ {\em Soft drinks}  & $\{*, *, Finistere\}$&$\rightarrow$ {\em Canned pork}   \\\hline
    $\{> 65, F, $ Mause$\}$& $ \rightarrow$ {\em Dairy}                & $\{*, *, Nord\}$&$\rightarrow$ {\em Imported beers} & $\{*, *, Cotes\ d'Armor\}$&$\rightarrow$ {\em Canned pork}   \\\hline
    $\{> 65, *, $ Aube$\}$& $ \rightarrow$ {\em Dairy}                 & $\{*, F, Nord\}$&$\rightarrow$ {\em Liquids}      & $\{> 65, F, Morbihan\}$&$\rightarrow$ {\em Fresh milk}   \\\hline
    $\{> 65, F, $ Haute Vienne$\}$& $ \rightarrow$ {\em Dairy}         & $\{*, F, Nord\}$&$\rightarrow$ {\em Beers}        & $\{*, *, Nord\}$&$\rightarrow$ {\em Beer}   \\\hline
    $\{> 65, F, $ Maine et Loire$\}$& $ \rightarrow$ {\em Dairy}       & $\{*, *, Finistere\}$&$\rightarrow$ {\em Butters} & $\{*, *, Nord\}$&$\rightarrow$ {\em Sparkling liquors}   \\\hline
    $\{> 65, *, $ Val de Marne$\}$& $ \rightarrow$ {\em Dairy}         & $\{*, F, Garonne\}$&$\rightarrow$ {\em Drugstore} & $\{*, *, Vienne\}$&$\rightarrow$ {\em Breakfast biscuits}   \\\hline
	\end{tabular}
  \caption{
    Top-10 demographics association rules, according to different interestingness measures.
    Rules are denoted \{age, gender, department\} $\rightarrow$ {\em product category}.
    Product categories were translated to English for clarity. French departments were left unchanged.
    Product brands were removed for confidentiality.
  }
  \label{tab:eyecatcher}
\end{table*}

The question we ask ourselves is {\bf how different are the rule rankings produced by
existing interestingness measures in the retail domain?}
To address this question, we examine the rankings produced by 34 measures~\cite{geng2006ACM,Lenca2007}.
This effort was conducted for three mining scenarios designed by experienced analysts from the marketing studies department of Intermarch\'e. In the first scenario, \demoassoc, the analyst provides a target product category and expects rules of the form {\em customer segment} $\rightarrow$ {\em category}, i.e. {\em customers who belong to the described segment purchase products in the target category.}
In the other two scenarios, the analyst provides a target product $p$ and expects rules of the form
{\em set of products} $\rightarrow$ $p$.
Such rules are either extracted based on a receipt-centric view,
where products are grouped by receipt (\prodassocreceipt scenario),
or based on a customer-centric view,
where products are grouped by customer across several receipts (\prodassocclient scenario).
Our first finding is that existing interestingness measures can be automatically grouped into 6 families of similar measures, regardless of the mining scenario.

We then conducted a user study with two experienced domain experts from Intermarch\'e in
order to address the following question: {\bf out of the 6 families of
  interestingness measures, which ones are meaningful?}
Our study lets analysts choose one of 3
mining scenarios along with target products or
categories. Analysts also choose an
interestingness measure without knowing which one it is. Their
interactions with the resulting list of association rules were
observed and their feedback recorded in a free-text form. Overall, ranking rules by decreasing confidence
was preferred.
Combined with the minimum support threshold used in the mining phase, this ranking promotes rules that are considered reliable.
However, the preference of the analysts changes when filters are available to narrow down the set of rules to specific product categories.
In this case, they favor the compromise between confidence and support offered, for instance, by the Piatetsky-Shapiro's measure~\cite{Piatetsky1991KDD}, as it promotes rules that are observed more frequently.


\project\ is made possible with \algo, our
distributed pattern mining algorithm that is able to mine millions of patterns in
a few minutes~\cite{jlcm}. \algo\ can be constrained to focus on different
customer demographics and product taxonomies.  Thus, in addition to
typical associations between products, it finds associations
between customer segments and products and between products and
 categories.

In summary, this paper presents \project, a joint effort between
researchers in Academia and business experts in Intermarch\'e.
\project\ is a framework that lets analysts compare and contrast
different interestingness measures (over 30 measures described
in~\cite{geng2006ACM}). The context and goals of the
work are provided in Section~\ref{sec:context}. The architecture
of \project is overviewed in Section~\ref{sec:arch}.
In Section~\ref{sec:xp:empirical}, \project is deployed to perform an empirical evaluation of interestingness measures and identify 6 groups of measures.
These groups are then evaluated by retail experts in Section~\ref{sec:xp:user}.
The related work is summarized in Section~\ref{sec:related}.
Planned and possible evolutions are finally discussed in Section~\ref{sec:evolutions}.

\section{Context} \label{sec:context}

\subsection{Dataset}
We represent a dataset $\cal D$ as a set of records of the form $\langle t, c, p \rangle$, where $t$ is a unique receipt identifier, $c$ is a customer, and $p$ is a product purchased by $c$.
When a customer purchases multiple products at the same time,
several records with the same receipt identifier $t$ are generated.
The set of receipt identifiers is denoted as $T$.
Each receipt identifier is associated with a unique customer, and multiple receipt identifiers can be associated with the same customer.
We do not use product price or product cardinality in this work.
The complete dataset contains over 290 million unique receipts, spanning 3.5 billion records, generated at a retail chain consisting of \num{1884} stores
over the whole year of 2013.
Table~\ref{tab:terms} summarizes notations and cardinalities of the data.

The set of customers, {\mf C}, contains over $9$ millions customers.
Each customer has demographic attributes.
In this study, we focus on 3 attributes: {\em age, gender} and {\em location}.
The attribute {\em age} takes values in \{{\em <35, 35-49, 50-65, >65}\} and the attribute {\em location} admits French departments as values.
Each customer segment is described by a set of user attribute values that are interpreted in the usual conjunctive manner.
For example, the segment $\{\mathit{<35},\mathit{Paris}\}$ refers to young Parisian customers.

We use $\mathit{demo(c)}$ to refer to the set of attribute values of a customer $c$.
For example, \{{\em 35-49, female, Calvados}\} represents a
48 year old {\em female} from the {\em Calvados} department,
whom we will refer to as {\em Mary}.

\begin{figure}[tb]
	\centering
	\begin{tikzpicture}[>=latex]
    \node (root) at (-1,0) {[all]};

    \node (A_farleft) at (-3, -1) {Vegetables};
    \draw [->] (root.south) -- (A_farleft.north east);

    \node (A_left) at (-1,-1) {Grocery};
    \draw [->] (root.south) -- (A_left.north);

    \node (A_right) at (1,-1) {Beverages};
    \draw [->] (root.south) -- (A_right.north west);

    \node (A_farright) at (3,-1) {...};
    \draw [->,dotted] (root.south) -- (A_farright.north west);

    \node (B_farleft) at (-3,-2) {Organic};
    \draw [->] (A_left.south) -- (B_farleft.north east);

    \node (B_left) at (-1, -2) {Chocolate bars};
    \draw [->] (A_left.south) -- (B_left.north);

    \node (B_right) at (1, -2) {Breakfast};
    \draw [->] (A_left.south) -- (B_right.north west);

    \node (B_farright) at (3, -2) {...};
    \draw [->,dotted] (A_left.south) -- (B_farright.north west);

    \node (C_farleft) at (-3, -3) {Stuffed};
    \draw [->] (B_left.south) -- (C_farleft.north east);

    \node (C_left) at (-1, -3) {Family sized};
    \draw [->] (B_left.south) -- (C_left.north);

    \node (C_right) at (1, -3) {Desserts};
    \draw [->] (B_left.south) -- (C_right.north west);

    \node (C_farright) at (3, -3) {...};
    \draw [->, dotted] (B_left.south) -- (C_farright.north west);

    \node (D_l) at (-2, -4) {[...]};
    \draw [->] (C_left.south) -- (D_l.north east);
    \node (D) at (-1, -4) {products};
    \draw [->] (C_left.south) -- (D.north);
    \node (D_r) at (0, -4) {[...]};
    \draw [->] (C_left.south) -- (D_r.north west);
	\end{tikzpicture}
	\caption{\label{fig:taxonomy}Extract from our products taxonomy.}
\end{figure}
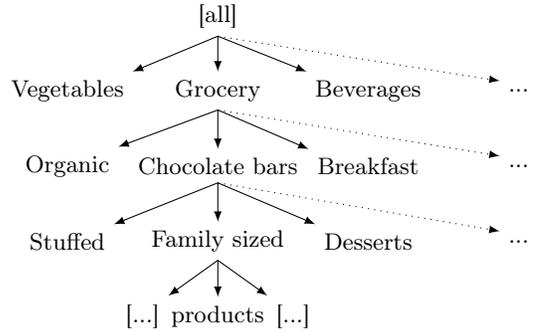

The set of products {\mf P} contains over \num{200000} entries, out of
which \num{55786} have been sold more than a thousand times.  Products
are organized in a taxonomy with \num{19557} nodes over 4 levels.
Figure~\ref{fig:taxonomy} shows a sample from our taxonomy.  Products
are leaf nodes, and belong to all their ancestor categories.
The set of categories a product $p$ belongs to is denoted
as $\mathit{cat(p)}$. For example, {\em chocolate cream} belongs to the categories
{\em Fresh food, Dairy, Ultra fresh} and {\em Desserts}.

\subsection{Mining Customer Receipts}

\subsubsection{Dataset Preparation\label{sec:dataprep}}
Our analysts are interested in studying two kinds of buying patterns: those
representing associations between customer segments and a product
category (e.g. {\em young people in the north of France consume sodas}),
and those associating a set of products to a single product (e.g.
{\em people who purchase pork sausage and mustard also buy dry Riesling}).
In all cases the analyst specifies ${\cal B}$,
the set of association rules' targets.

In the first case, coined \demoassoc,
${\cal B}$ contains one or more categories.
The analyst expects rules of the form {\em customer segment} $\rightarrow$ {\em category}, i.e. customers who purchase products in the target category.
The second case comes in two variants:
\prodassocreceipt, a receipt-centric view where products are found in the same receipt, and \prodassocclient, a customer-centric view where products are purchased by the same customer over time.
In these variants, ${\cal B}$ only contains products (as opposed to categories in the first senario)
and the analyst expects rules of the form {\em set of products} $\rightarrow$ {\em target product} $p \in {\cal B}$.

The dataset \mf D is transformed into a collection of transactions \mf T that is given as input to the mining process, as summarized in Table~\ref{tab:scenarios}.
The set \mf T is constructed differently for each scenario.

\begin{table}[tb]
\centering
\begin{tabular}{|c|c|c|}
\hline
{\bf Term} & {\bf Description} & {\bf Cardinalities}\\
\hline
$\mf D$  & Raw records $\langle t,c,p\rangle$            & \num{3502834638}  \\
\hline
$T$      & Set of receipts                                & \num{290734163}\\
\hline
$\mf C$  & Set of customers                              & \num{9267961} \\
\hline
$\mf P$  & Set of products                               & \num{222228}\\
\hline
$\mf T$  & Set of transactions                           & \demoassoc: \num{9267961}\\
         &                                               & \prodassocreceipt: \num{290734163} \\
         &                                               & \prodassocclient: \num{9267961} \\
\hline
\end{tabular}
\caption{Notations and cardinalities \label{tab:terms}}
\end{table}

In \demoassoc, a transaction is a tuple built for each receipt $\langle t, c, p \rangle$ by associating $\mathit{demo(c)}$ with $\mathit{cat(p)}$.
For example, for \emph{Mary},
the record $\langle 234567, \mathit{Mary, chocolate\ cream}\rangle$ is mapped to the transaction
$\langle$ 35-49, \textit{female, Calvados, chocolate\ cream, Fresh food, Dairy, Ultra fresh, Desserts}$\rangle$.
Thus, the number of transactions is equal to \mf{|D|}, and each transaction contains both,
the segments a customer belongs to, and the categories of the product purchased.

In \prodassocreceipt, \mf T is built by grouping the records in \mf D by receipt identifier, $t$.
Hence, for each $t$, we generate a transaction as the set of products bought in a single visit to the store $\{p|\langle t, c, p \rangle\in\mf D\}$.
For example, if our user {\em Mary} has a store receipt containing the products {\em cream, yoghurt, cola}, a transaction containing the 3 products is generated.
This leads to a total of $|T|$ transactions, where each transaction is a subset of the set of products, \mf P.

In \prodassocclient, we generate the set of transactions \mf T by grouping records in \mf D by customer.
For each customer $c$ , we generate a single transaction containing all products ever purchased by her $\{p|\langle t, c, p \rangle\in\mf D\}$.
We obtain \mf{|C|} transactions, each of which is a subset of \mf P.
This use case enables the discovery of patterns occurring over several visits to a store.

Table~\ref{tab:terms} contains the number of transactions in each scenario.
In \prodassocclient, the number of transactions is less important than in \prodassocreceipt,
but each transaction contains more products: 214, on average, whereas the average receipt contains 12 products.


\subsubsection{Mining Scenarios\label{sec:contextmining}}

Given a frequency threshold $\varepsilon \in [1,n]$, an itemset $P$ is
said to be {\em frequent} in a transactions set ${\cal T}$ iff
$\mathit{support}_{\cal T}(P) \geq \varepsilon$ where
$\mathit{support}_{\cal T}(P)$ is the number of transactions in $\cal
T$ that contain all items in $P$. As indicated in
Table~\ref{tab:scenarios}, we set the frequency threshold to different
values in different scenarios because they differ in the cardinalities of their transactions.  Moreover, because marketing actions are decided and applied nation-wide,
they are expected to concern at least \num{1000} customers,
and preferably more than \num{10000}.

\begin{table*}[th!]
	\centering
	\begin{tabular}{|c|c|c|}
		\hline
		\textbf{Target Associations}        & \textbf{Input transactions ${\cal T}$} & \textbf{Desired association rules}\\ \hline
		\demoassoc: & $ \{ \mathit{demo}(c) \cup \mathit{cat}(p) | \langle t, c, p \rangle\in\mf D\} $   &  A segment tends to purchase products in a category. \\
		{\em segment} $\rightarrow$ {\em category}  &  min support is \num{1000}  &  $\{< 35, F, *\}\rightarrow$ {\em Baby food}  \\
	  &                                                                   &  $\{*, *, Nord\}\rightarrow$ {\em Sodas} \\
	  &                                                                   &  $\{> 65, *, Gironde\}\rightarrow$ {\em Bordeaux wine} \\
			\hline
	 	\prodassocreceipt:& \(\displaystyle  \{ \cup_{\langle t, c_j, p_i \rangle\in\mf D} p_i | t\in T\} \)   & Products purchased \emph{simultaneously}. \\
	  {\em product(s)} $\rightarrow$ {\em product}  &       min support is \num{1000}      & \{{\em vanilla cream}\}$\rightarrow$ {\em chocolate cream}\\
		\hline
		\prodassocclient: & \(\displaystyle  \{ \cup_{\langle t_j, c, p_i \rangle\in\mf D} p_i | c\in\mf C\} \)  &  Customers' product associations over time. \\
		{\em product(s)} $\rightarrow$ {\em product} 	&       min support is \num{10000}     & \{{\em Pork sausage, mustard\}}$\rightarrow$ {\em dry Riesling} \\
		\hline
	\end{tabular}
\caption{Our mining scenarios and example association rules.}
\label{tab:scenarios}
\end{table*}

\colorlet{g1aColor}{rgb:yellow,1;white,2}
\colorlet{g1bColor}{rgb:yellow,1;white,2}

\colorlet{g2Color}{rgb:red,1;white,2}
\colorlet{g3Color}{rgb:green,1;white,2}
\colorlet{g5Color}{rgb:gray,1;white,2}
\colorlet{g6Color}{rgb:blue,1;white,2}
\colorlet{g4Color}{rgb:brown,1;white,2}
\begin{table*}[th!]
{\small
  \centering
  \begin{tabular}{|l|l|c|}
		\hline
    \textbf{Measure}  &   \textbf{Formula} &\textbf{Group}\\\specialrule{.15em}{0em}{0em}
    One-Way Support                  & $P(B|A)\times log_{2}\frac{P(AB)}{P(A)P(B)}$ &\cellcolor{g1aColor}\\\cline{1-2}
    Relative Risk                    & $P(B|A) / P(B|\neg A)$ &\cellcolor{g1aColor}\\\cline{1-2}
    Odd Multiplier                   & $\frac{P(AB)P(\neg B)}{P(B)P(A \neg B)}$ &\cellcolor{g1aColor}\\\cline{1-2}
    Zhang                            & $\frac{P(AB)-P(A)P(B)}{max(P(AB)P(\neg B), P(B)P(A \neg B))}$ &\cellcolor{g1aColor}\\\cline{1-2}
    Yule's Q $\Diamond$                        & $\frac{P(AB)P(\neg A \neg B)-P(A \neg B)P(B \neg A)}{P(AB)P(\neg A \neg B)+P(A \neg B)P(B \neg A)}$ &\cellcolor{g1aColor}\\\cline{1-2}
    Yule's Y $\Diamond$                        & $\frac{\sqrt{P(AB)P(\neg A \neg B)}-\sqrt{P(A \neg B)P(B \neg A)}}{\sqrt{P(AB)P(\neg A \neg B)}+\sqrt{P(A \neg B)P(B \neg A)}}$ &\cellcolor{g1aColor}\\\cline{1-2}
    Odds Ratio $\Diamond$                      & $\frac{P(AB)P(\neg A \neg B)}{P(A \neg B)P(B \neg A)} $ &\cellcolor{g1aColor}\\\cline{1-2}
    Information Gain $\ast$$\ominus$          & $\mathit{log}(P(AB)/(P(A)P(B)))$ &\cellcolor{g1aColor}\\\cline{1-2}
    Lift $\ast$$\ominus$                            & $P(AB)/(P(A)P(B))$&\cellcolor{g1aColor}\multirow{-8}{*}{$G_1^a$}\\\specialrule{.12em}{0em}{0em}
    Added Value $\ast$                     & $P(B|A) - P(B)$ &\cellcolor{g1bColor}\\\cline{1-2}
    Certainty Factor $\ast$                & $(P(B|A) - P(B)) / (1 - P(B))$ &\cellcolor{g1bColor}\\\cline{1-2}
    Confidence / Precision $\ast$$\otimes$          & $P(B|A)$ &\cellcolor{g1bColor}\\\cline{1-2}
    Laplace Correction $\ast$$\otimes$          & $\frac{\mathit{support}(AB)+1}{\mathit{support}(A)+2}$   &\cellcolor{g1bColor}\\\cline{1-2}
    Loevinger $\dagger$                       & $1 - \frac{P(A)P(\neg B)}{P(A \neg B)}$ &\cellcolor{g1bColor}\\\cline{1-2}
    Conviction $\dagger$                      & $\frac{P(A)P(\neg B)}{P(A \neg B)}$ &\cellcolor{g1bColor}\\\cline{1-2}
    Example and Counter-example Rate & $1 - \frac{P(A \neg B)}{P(AB)}$ &\cellcolor{g1bColor}\\\cline{1-2}
    Sebag-Schoenauer                 & $\frac{P(AB)}{P(A \neg B)}$ &\cellcolor{g1bColor}\\\cline{1-2}
    Leverage                         & $P(B|A) - P(A)P(B)$ &\cellcolor{g1bColor}\multirow{-8}{*}{$G_1^b$}\\\specialrule{.15em}{0em}{0em}
    Least Contradiction              & $\frac{P(AB)-P(A \neg B)}{P(B)}$ &\cellcolor{g2Color}\\\cline{1-2}
    Accuracy                         & $P(AB) + P(\neg A \neg B)$ &\cellcolor{g2Color}\multirow{-2}{*}{$G_2$}\\\specialrule{.15em}{0em}{0em}
    Pearson's $\chi^2$ $\triangleright$              & $|{\cal T}| \times \left(\frac{(P(AB)-P(A)P(B))^2}{P(A)P(B)} + \frac{(P(\neg A B)-P(\neg A)P(B))^2}{P(\neg A)P(B)}\right)$&\cellcolor{g3Color}\\
&$+ |{\cal T}| \times \left(\frac{(P(A \neg B)-P(A)P(\neg B))^2}{P(A)P(B)} + \frac{(P(\neg A \neg B)-P(\neg A)P(\neg B))^2}{P(\neg A)P(\neg B)}\right)$&\cellcolor{g3Color}\\\cline{1-2}
    Gini Index $\triangleright$                      & $P(A)\times(P(B|A)^2 + P(\neg B|A)^2) + P(\neg A)\times(P(B|\neg A)^2+$&\cellcolor{g3Color}\\
&$P(\neg B|\neg A)^2)-P(B)^2 - P(\neg B)^2$&\cellcolor{g3Color}\\\cline{1-2}
    J-measure                        & $P(AB)log(\frac{P(B|A)}{P(B)})+P(A \neg B)log(\frac{P(\neg B|A)}{P(\neg B)})$ &\cellcolor{g3Color}\\\cline{1-2}
    $\Phi$ Linear Correlation Coefficient & $\frac{P(AB)-P(A)P(B)}{\sqrt{P(A)P(B)P(\neg A)P(\neg B)}}$&\cellcolor{g3Color}\\\cline{1-2}
    Two-Way Support Variation        & $P(AB)\times log_{2}\frac{P(AB)}{P(A)P(B)} + P(A \neg B)\times log_{2}\frac{P(A \neg B)}{P(A)P(\neg B)} +$ &\cellcolor{g3Color}\\
& $P(\neg A B)\times log_{2}\frac{P(\neg A B)}{P(\neg A)P(B)} + P(\neg A \neg B)\times log_{2}\frac{P(\neg A \neg B)}{P(\neg A)P(\neg B)}$&\cellcolor{g3Color}\\\cline{1-2}
    Fisher's exact test               & $\frac{\binom{|{\cal T}|\times P(B)}{|{\cal T}| \times P(AB)}\binom{|{\cal T}| \times P(\neg B)}{|{\cal T}| \times P(A\neg B)}}{\binom{|{\cal T}|}{|{\cal T}|\times P(A)}}$ &\cellcolor{g3Color}\\\cline{1-2}
    Jaccard                          & $P(AB) / (P(A)+P(B)-P(AB))$ &\multirow{-10}{*}{$G_3$}\cellcolor{g3Color}\\\specialrule{.15em}{0em}{0em}
    Cosine                           & $\frac{P(AB)}{\sqrt{P(A)P(B)}}$ &\cellcolor{g4Color}\\\cline{1-2}
    Two-Way Support                  & $P(AB)\times log_{2}\frac{P(AB)}{P(A)P(B)}$&\cellcolor{g4Color}\multirow{-2}{*}{$G_4$}\\\specialrule{.15em}{0em}{0em}
    Piatetsky-Shapiro                & $P(AB)-P(A)P(B)$ &\cellcolor{g5Color}\\\cline{1-2}
    Klosgen                          & $\sqrt{P(AB)}\mathit{max}(P(B|A)-P(B), P(A|B)-P(A))$ &\cellcolor{g5Color}\\\cline{1-2}
    Specificity                      & $P(\neg B | \neg A)$ &\cellcolor{g5Color}\multirow{-3}{*}{$G_5$}\\\specialrule{.15em}{0em}{0em}
    Recall                           & $P(A|B)$ &\cellcolor{g6Color}\\\cline{1-2}
    Collective Strength              & $\frac{P(AB)+P(\neg B|\neg A)}{P(A)P(B)+P(\neg A)P(\neg B)}\times\frac{1-P(A)P(B)-P(\neg A)P(\neg B)}{1 - P(AB) - P(\neg B|\neg A)}$&\cellcolor{g6Color}\multirow{-2}{*}{$G_6$}\\\hline
  \end{tabular}
  \caption{Interestingness measures of a rule $A \rightarrow B$.
    $\ast$, $\triangleright$ indicate measures that produce the same rule ranking when a single target is selected.
    $\Diamond$, $\dagger$, $\ominus$, $\otimes$ indicate measures that always produce the same rule ranking.
      $|{\cal T}|$ is the number of transactions. $P(A) = support(A)/|{\cal T}|$.
  }
  \label{tab:measures}
}
\end{table*}

\begin{figure*}[th!]
	\centering
  \includegraphics[width=\textwidth]{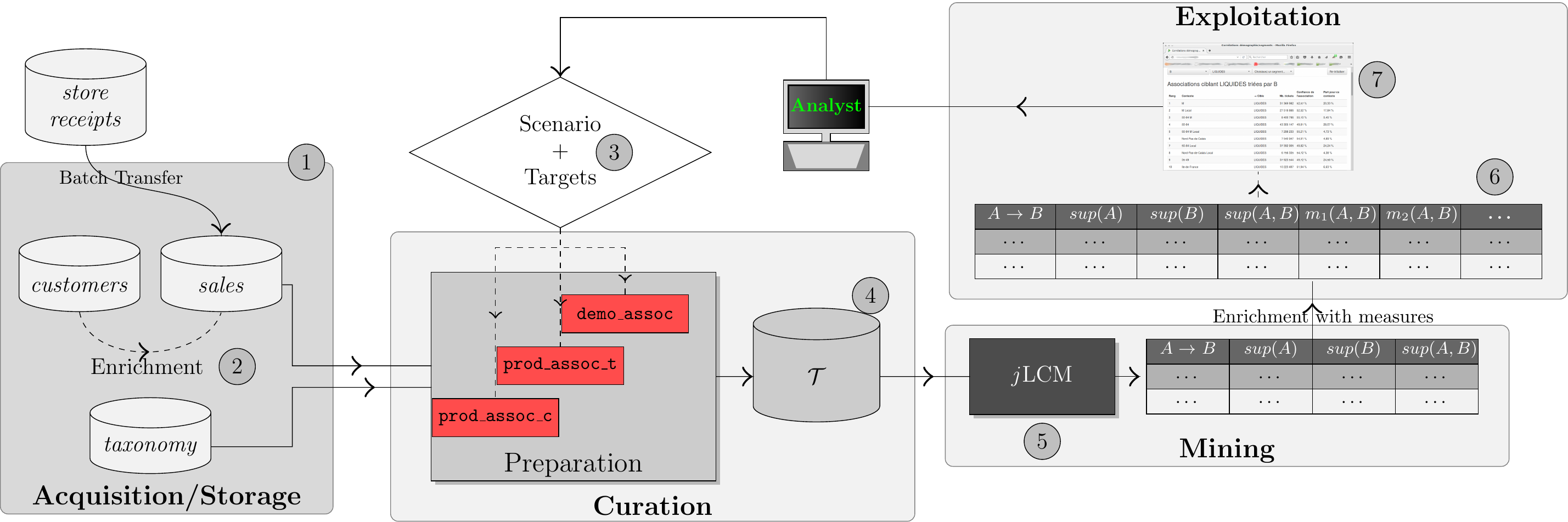}
	\caption{Overview of the architecture\label{fig:archoverview}}
\end{figure*}

An itemset $P$ is
said to be closed iff there exists no itemset $P'\supset P$ such that
$\mathit{support}_{\cal T}(P) = \mathit{support}_{\cal
T}(P')$~\cite{DBLP:conf/icdt/PasquierBTL99}.  The number of closed frequent itemsets
can be orders of magnitude less important than the number of frequent itemsets, while
providing the same amount of information on ${\cal T}$.  Several
algorithms, including ours, focus on extracting frequent closed itemsets,
increasing performance and avoiding redundancy in
results~\cite{conf/dmkd/PeiHM00,DBLP:conf/fimi/UnoKA04}.

We consider our mining scenarios described in Section~\ref{sec:dataprep}. Each scenario leads to the
construction of a collection of transactions $\cal T$, where a
transaction is a set of items. Given ${\cal T}$, a frequency threshold $\varepsilon$, we find all
closed frequent itemsets, and use them to derive association
rules~\cite{datamining}.  Each itemset $P$ implies an
association rule of the form $A\rightarrow B$ where $A,B$ is a
partition of $P$.  $A$ is the antecedent of the rule, and $B$ its
consequent.
In \demoassoc, $A$ is a customer segment and $B$ is a single product category.
In \prodassocreceipt and \prodassocclient, $A$ is a set of
products ($A\subseteq \cal P$) and $B$ is a single product.
Analysts generally focus on particular products or
product categories.
This is why they specify the list of targets $\cal B$ in each scenario.
Table~\ref{tab:scenarios} contains example association rules
extracted from our dataset, for each scenario.

\subsection{Interestingness Measures\label{sec:measures}} Large
datasets often contain millions of frequent closed itemsets, and each
of them may lead to several association rules.  The ability to identify
valuable association rules is therefore of the utmost
importance to avoid drowning analysts in useless information.
Association rules $A\rightarrow B$ were originally selected using
thresholds for support ($\mathit{support}_{\cal T}(A\cup B)$) and
confidence ($\frac{\mathit{support}_{\cal T}(A\cup
B)}{\mathit{support}_{\cal T}(A)}$)~\cite{AggrawalSIGMOD1993}.
However
using two separate values, and guessing the right threshold is not natural.
Furthermore, support and confidence do not always coincide with the interest of analysts.
Hence, a number of interestingness measures that serve different analyses needs were proposed in the literature~\cite{geng2006ACM,Minato2014KDD,LiuICDE2011}.
Table~\ref{tab:measures} summarizes the interestingness measures we use in this work.
The first column contains the name of the measure, the second its expression.
The last column will be referred to later.

\subsection{Goal}
\label{sec:goal}
Our goal is to help analysts test and compare
different interestingness measures on association rules extracted from from $\cal D$.
An analyst can specify one of 3 mining scenarios, \demoassoc, \prodassocreceipt, and \prodassocclient, and
one or several targets (categories in the case of \demoassoc, products
in the case of the other two), and \project generates a ranked list
of association rules sorted using different
interestingness measures.

\pagebreak


\section{Architecture}
\label{sec:arch}

Figure~\ref{fig:archoverview} contains the main components of \project\ and their interactions.
The first module is \textbf{acquisition and storage}.
Sales records are produced locally at each store, and are loaded  daily into a data center \circled{1}.
Records are stored in a \textit{sales} table, and are augmented with customer segments coming from the \textit{customers} table \circled{2}.
\project's \textbf{curation} module is used to build transactions.
The analysts selects a mining scenario and a set of input targets \circled{3}, which are used to generate the appropriate collection of transactions \mf T \circled{4}. 
\project's \textbf{mining} component relies on \algo, an open-source pattern mining library that we developed~\cite{jlcm}, to compute a set of association rules matching the input targets \circled{5}.
\project's \textbf{exploitation} component computes the quality of produced rules according to each interestingness measure \circled{6}, and loads them into a database.
Results are presented to the analyst through a web application \circled{7}.
We now describe the details of each component of \project.

\subsection{Acquisition and storage\label{sec:archstore}}

Each of the \num{1884} stores locally maintains a log of all customer
transactions completed during the day.  Whenever a customer checks
out, a receipt is generated, indicating the list of products
purchased, their price, as well as potential discounts.  
These receipts are logged under the form of $\langle r, c, p \rangle$
triples and stored in a write-ahead log.  Once a day, during the
store's closing time, this log is transmitted to the main data center
that centralizes all sales records.

We rely on Hadoop YARN~\cite{yarn} to administer the cluster dedicated
to storing sales records.  All data is stored in an HBase
database~\cite{hbase}, and processing is performed using the Hadoop
MapReduce framework~\cite{mapreduce}.  Sales records are stored in the
\textit{sales} table. To avoid redundancy and ease data processing,
records are grouped by receipt before being stored in \textit{sales}.
Thus, each receipt is a line in the table, and the content of the
receipt is stored in the \textit{meta} column family.  We leverage
HBase's flexibility on columns by recording each product identifier as a
column qualifier, with information such as the cardinality and the
unit-price as a value.  The row key of each receipt  is defined
as \textit{storeId-day-customerId-receiptId}.  The \textit{sales} table
is configured to be sorted by row key.  This allows operations such as
selecting the sales records of a given store to be efficiently
performed in a single scan, while selecting a specific time
period can also be done by combining \num{1884} ranges (one per store
identifier).  Given that customer purchases may vary significantly between geographical areas~\cite{dice15} and over time, these two
operations are frequently used by analysts.  This data layout is
optimized to perform these selections efficiently, without incurring unnecessary
reads. That allows to store large amounts of data without increasing the cost of analyzing a fixed
number of records.  Sales logs transferred from the stores are
initially stored on the distributed file system HDFS, and then loaded
into HBase using MapReduce, as a daily batch job.

Each  customer
constitutes an entry in the \textit{customers} table, which records
the segments she belongs to.  
After loading the sales
records into the database, we enrich the \textit{sales} table using
another MapReduce job.  For each new record, 
the receipt is augmented with the user segments by
querying the \textit{customers} table and copying these segments to
the \textit{meta} column family in \textit{sales}.  Hence, each sales
record is assigned a snapshot of the user information at the time the
receipt was generated.

\subsection{Curation}
\comment{
\begin{figure*}[t!]
	\centering
	\includestandalone[width=0.90\textwidth]{figures_latex/hbaseTable}
	\caption{\textit{sales} table in HBase\label{fig:salesTable}}
\end{figure*}
}

As described in Section~\ref{sec:dataprep}, mining customer receipts
begins with the construction of a transactions dataset $\cal T$
following the requirements of the analyst.  This operation is
performed using MapReduce jobs executed on the \textit{sales} table.
In the case of \demoassoc, a single {\em map} operation is
sufficient.  The product taxonomy is loaded in memory by all mappers
through the distributed cache, and, given a row, for each product
registered in the \textit{products} column family, a transaction containing
its categories is generated.  Customer segments are directly available in
the \textit{meta} column family thanks to the enrichment phase and are
added to the transactions.
As described in Section~\ref{sec:archstore},
records are already grouped by receipt when stored in \textit{sales},
thus no further processing is necessary for an analysis in \prodassocreceipt.
Each line of \textit{sales} generates one transaction containing the set of products.
In \prodassocclient, the products bought by a given
customer are grouped using a reduce operation with the customer
identifier as a key to generate a transaction.
In all cases, at the end of this phase the dataset $\cal T$ is stored on HDFS as a text file,
with one line per transaction.

\begin{algorithm}[t!]
	\begin{small}
		\caption{Extracting itemsets with \algo\label{alg:LCMMR}}
		\KwData{dataset $\cal T$, minimum support threshold $\varepsilon$, target items $\cal B$}
		\KwResult{Output all closed itemsets in ${\cal T}$ containing an item from $\cal B$}
		\SetKwProg{myproc}{Function}{}{}
		\myproc{map$(E \in {\cal T})$\label{line:mapstart}}{
			\ForEach{$b \in {\cal B}$}{
				\If{$b \in E$}{
					\textbf{output} $(b,E)$\label{line:mapend}\\
				}
			}
		}
		\vspace{0.1 cm}
		\SetKwProg{myproc}{Function}{}{}
		\myproc{reduce$(b\in {\cal B},{\cal T}_{\{b\}}\subseteq {\cal T},\varepsilon)$\label{line:reducestart}}{
			\KwData{target $b$, filtered dataset ${\cal T}_{\{b\}}$, freq. threshold $\varepsilon$}
			\KwResult{Output all closed itemsets containing $e$}
			\algo$(\emptyset,b,{\cal T}_{\{b\}},\varepsilon)$\label{line:reduceend}\\
		}
		\vspace{0.1 cm}
	       \SetKwProg{myproc}{Function}{}{}
                \myproc{\algo$(P,e,{\cal T}_P,\varepsilon)$\label{line:jlcmstart}}{
                        \KwData{Base itemset $P$, extension item $e$, supporting transactions $T_P$, freq. threshold $\varepsilon$}
                        \KwResult{Output all closed itemsets containing $\{e\}\cup P$}
                                $Q \gets clo(\{e\}\cup P)$  \tcp*[f]{Closure computation}  \label{line:LCMitemsetClosure}\\
                                \If(\tcp*[f]{Unicity check}){$\mathit{max(Q \setminus P)} = e$}{ \label{line:LCM1stparent}
                                        \If{$|Q| \geq 2$}{  \label{line:LCMlengthCheck}
                                                \textbf{output} $(Q,\mathit{support}_{{\cal T}_P}(Q))$ \label{line:LCMoutput}
                                        }
                                        \ForEach(){$i \in \mathit{freq}_\varepsilon({\cal T}_Q) \mid i < e$}{\label{line:LCMaugmentations} 
                                                \algo$(Q,i,{\cal T}_Q,\varepsilon)$    \label{line:LCMrecCall}
                                        }
                                }
                }
	\end{small}
\end{algorithm}

\begin{figure}

	\begin{tikzpicture}[>=latex]
    \node(table) at (-1,-0.4) {${\cal T}_e=\{\{a,b,c,x,y\}, \{a,c,y\}, \{a,b,x,y\}, \{b,c,x,y\}\}$};

    \node (x) at (-2.5,0) {$\{x\},4$};
    \node (y) at (0.5,0) {$\{y\},3$};
    \node (ax)  at (-4,1.2) {$\{a,b,x\},2$};
    \node (bx)  [draw] at (-3.5,2.3) {$\{b,x\},3$};
    \node (abx)  [draw] at (-3.5,3.5) {$\{a,b,x\},2$};
    \node (cx)  [draw] at (-2,1.2) {$\{b,c,x\},2$};
    \node (ay)  [draw] at (-0.4,1.2) {$\{a,y\},3$};
    \node (by)  [draw] at (0.5,2.3) {$\{b,y\},3$};
    \node (aby)  [draw] at (0.5,3.5) {$\{a,b,y\},2$};
    \node (cy)  [draw] at (2.0,1.2) {$\{c,y\},3$};
    \node (acy)  [draw] at (2.0,2.9) {$\{b,c,y\},2$};
    \draw [->,thick] (x.north)--(ax.south) node[above, midway, xshift=-0.6cm, yshift=-0.4cm] {$\langle\{x\}, a\rangle$};
    \draw [->,thick] (x.north)--(bx.south) node[above, midway, xshift=0.3cm, yshift=0.4cm] {$\langle\{x\}, b\rangle$};
    \draw [->,thick] (x.north)--(cx.south) node[above, midway, xshift=0.5cm, yshift=-0.5cm] {$\langle\{x\}, c\rangle$};
    \draw [->,thick] (bx.north)--(abx.south) node[above, midway, xshift=-0.8cm, yshift=-0.3cm] {$\langle\{b,x\}, a\rangle$};
    \draw [->,thick] (y.north)--(ay.south) node[above, midway, xshift=-0.6cm, yshift=-0.4cm] {$\langle\{y\}, a\rangle$};
    \draw [->,thick] (y.north)--(by.south) node[above, xshift=0.6cm, yshift=-0.6cm] {$\langle\{y\}, b\rangle$};
    \draw [->,thick] (by.north)--(aby.south) node[above, midway, xshift=-0.8cm, yshift=-0.3cm] {$\langle\{b,y\}, a\rangle$};
    \draw [->,thick] (y.north)--(cy.south) node[above, midway, xshift=0.6cm, yshift=-0.5cm] {$\langle\{y\}, c\rangle$};
    \draw [->,thick] (cy.north)--(acy.south) node[above, midway, xshift=0.7cm, yshift=-0.2cm] {$\langle\{c,y\}, b\rangle$};
  \end{tikzpicture}

\caption{\algo enumeration trees over an example dataset ${\cal T}_e$, with $\varepsilon=2$ and ${\cal B}=\{x,y\}$.
An edge $\langle P,e\rangle$ represents an invocation of \algo, a node is a closed itemset $Q$ (Algorithm~\ref{alg:LCMMR}, Line~\ref{line:LCMitemsetClosure}).
Only boxed nodes are returned.
\label{fig:lcmtree}
}
\end{figure}
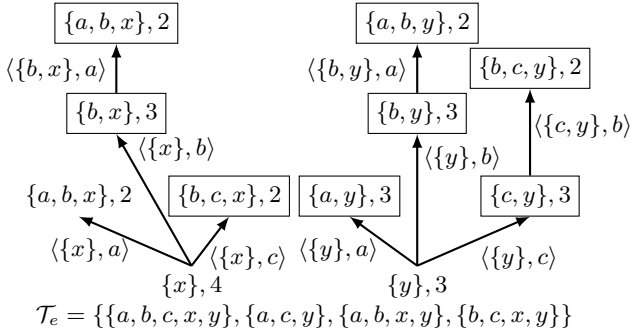

\pagebreak

\subsection{Mining}
\subsubsection{Extracting itemsets using \algo}

Generating association rules, presented in Section~\ref{sec:contextmining},
firstly requires the extraction of frequent itemsets from $\cal T$.
We rely on \algo, our Java implementation of the LCM
algorithm~\cite{DBLP:conf/fimi/UnoKA04} available as an open-source
library~\cite{jlcm}.

\algo is integrated in a MapReduce job, as detailed in Algorithm~\ref{alg:LCMMR}.
The itemset extraction job scans the input $\cal T$ once in the $\mathit{map}$ function, and builds for each target item $b$ in ${\cal B}$, a filtered dataset limited to transactions containing $b$: ${\cal T}_{\{b\}} = \{E\in {\cal T}, b\in E\}$ (Lines~\ref{line:mapstart}--\ref{line:mapend}).
This is done using the target items from ${\cal B}$ as intermediate keys for the $\mathit{reduce}$ function.
For each target, $\mathit{reduce}$ executes \algo on its filtered dataset (Lines~\ref{line:reducestart}--\ref{line:reduceend}).
\algo is a recursive algorithm that enumerates itemsets and computes their frequency following a depth-first tree-shaped traversal (Lines~\ref{line:jlcmstart}--\ref{line:LCMrecCall}).

As proposed by Uno {\textit et al.}~\cite{DBLP:conf/fimi/UnoKA04}, Line~\ref{line:LCMitemsetClosure} ensures that only closed itemsets are enumerated, and Line~\ref{line:LCM1stparent} avoids duplicating enumerations.
Closed itemsets are returned along with their support (Line~\ref{line:LCMoutput}), with the exception of singletons that cannot produce association rules.
In \demoassoc, itemsets should contain a single category only, so all categories except the target one are removed from transactions.

Figure~\ref{fig:lcmtree} depicts an example of \algo execution with two targets ${\cal B}=\{x,y\}$.
This leads to 2 separate enumeration trees, explored in parallel by two reducers.
In the case of the target $x$, on the left, \algo starts with the singleton itemset $\{x\}$ and recursively builds larger itemsets.
The closed itemset $\{a,b,x\}$ is encountered several times throughout the enumeration, but is only outputted once thanks to the test of Line~\ref{line:LCM1stparent}.
This technique allows us quickly obtain itemsets that satisfy our constraint,
{\em i.e.} all itemsets contain one of the targets from $\cal B$.
The job's run-time is dominated by the scan of $\cal T$ in the {\em map} phase,
which can be accelerated by the addition of worker nodes.
On average, each {\em reduce} task completes in 10 seconds.

\subsubsection{Evaluating relevant association rules}
Analysts aim at uncovering interesting association rules expressed as $A \rightarrow B$.
Evaluating the interestingness of an association rule
requires computing the support of itemsets $A$, $B$ and $A\cup B$ in
$\cal T$.  The standard method for mining association rules consists
in finding all frequent itemsets in the dataset, and then generating
the rules.  Given that our analysts have specified a restricted set
of targets ${\cal B}$, this approach would be wasteful.
This motivates our distribution of the itemsets extraction,
presented in the previous sub-section.
Our itemsets extraction job
gives the support of $B$ and $A\cup B$ for all association
rules we are interested in
({\em ie.} all $B$ satisfy $B = \{e\}, e \in {\cal B}$).
This job also materializes, as a prefix tree in a side-output file,
the set $\cal A$ of all antecedent itemsets,
whose support needs to be evaluated.

An second MapReduce job completes the evaluation of association rules.
Each {\em map} operation reads a transaction of ${\cal T}$ and counts the support all association rules' antecedents.
The {\em reduce} phase uses itemsets in $\cal A$ as intermediate keys and sums partial counts to obtain the total support.
This two-step approach avoids the computation of many itemsets that never appear as a rule antecedent.

\subsection{Exploitation}
\label{sec:exploitation}
The quality measures selected require at most $P(A)$, $P(B)$ and $P(A \cup B)$ to be computed, because,
given $|{\cal T}|$, other probabilities like $P(B|A)$ or $P(A \neg B)$ can be derived from them.
Therefore, we denormalize the results of the mining phase in order to store those 3 probabilities with each $A$ and $B$.
The supports of all rules' antecedents (providing $P(A)$) are centralized
and joined to the results of \algo (which provides $P(AB)$ and $P(B)$).
After this denormalization, each row represents an association rule and has enough information to compute its score.
This table is then augmented with 34 columns,
one for each measure implemented in \project, and listed in Table~\ref{tab:measures}.
Because large numbers are involved,
for {\em Fisher's exact test} we actually use the logarithm of the binomial coefficients,
which are computed as logarithms of the gamma function.
This makes the calculation feasible, but requires long iterations
so we do it in parallel again (this is easy to implement thanks to the denormalization).
The complete table is stored in a relational database.

The final component of \project is a web application allowing the analyst to explore this augmented table.
In any scenario,
the analyst picks a measure and selects a target product or category, or a set of target products or categories.
Association rules are then returned in a table and sorted according to the selected measure.

A rule like $\mathit{yoghurt}\rightarrow\mathit{cheese}$ is displayed with 3 values:
\emph{support} (number of customers who bought both cheese and yoghurt),
\emph{confidence} (fraction of yoghurt buyers who also bought cheese),
\emph{recall} (fraction of cheese buyers who also bought yoghurt).
During the user study these figures help the analyst quickly judge the volume of sales for each rule.

\section{Empirical evaluation}
\label{sec:xp:empirical}
We present an empirical evaluation of the 34 measures for association rules introduced in Section~\ref{sec:measures}.
Recall that our goal, stated in Section~\ref{sec:goal}, is to assist the analyst in selecting measures.
Our evaluation consists in comparing rankings produced by these measures on retail data to discover which measures differ significantly in practice.
We then use that similarity to classify ranking measures into \emph{groups}.
We annotate these groups based on the properties common to the group.
We discuss key insights obtained from rigorous experimentation on each group.
The goal of this evaluation is to automatically detect similarities between interestingness measures and reduce the number of candidate measures to present to analysts the user study (Section~\ref{sec:xp:user}).

We first present in Section~\ref{sec:correlations} methods used to compare ranked list.
Then, we compare measures in two different cases: rules having the same target (Section~\ref{sec:identicaltargets}) and rules having different targets (Section~\ref{sec:differenttargets}).
We conclude the empirical evaluation with the selection of representative measures in Section~\ref{sec:selection}.

\subsection{Ranking similarity measures}
\label{sec:correlations}
In this section, we discuss some methods for comparison of ranked lists.
The first three methods are taken from the literature.
We then introduce {\em NDCC}, a new parameter-free ranking similarity designed to emphasize differences at the top of the ranking.

We are given of a set of association rules $\mf R$ to rank.
We interpret each measure, $m$, as a function that receives a rule and generates a score, $m:{\mf R}\rightarrow \mathbb{R}$.
We use $L_{\mf R}^m$ to denote an ordered list composed of rules in $\mf R$, sorted by decreasing score.
Thus, $L_{\mf R}^m=<r_{1},r_{2},\ldots>$ s.t. $\forall i>i'\ m(r_{i})<m(r_{i'})$.
We generate multiple lists, one for each measure $m$, from the same set $\mf R$.
$L_{\mf R}^m$ denotes a ranked list of association rules according to measure $m$ where the rank of rule $r$ is given as $rank(r,L_{\mf R}^m)=|\{r'|r'\in \mf R,\ m(r')\geq m(r)\}|$.
To assess the dissimilarity between two measures, $m$ and $m'$, we compute the dissimilarity between their ranked lists, $L_{\mf R}^m$ and $L_{\mf R}^{m'}$.
We use $r^m$ as a shorthand notation for $rank(r,L_{\mf R}^m)$.

\subsubsection{Spearman's rank correlation coefficient}

Given two ranked lists $L_{\mf R}^m$ and $L_{\mf R}^{m'}$, {\em Spearman's rank correlation}~\cite{daniel1978applied}
computes a linear correlation coefficient that varies between $1$ (identical lists) and $-1$ (opposite rankings) as shown below.
$$\mathit{Spearman(L_{\mf R}^m,L_{\mf R}^{m'})} = 1 - \frac{6\sum\limits_{r\in \mf R}{(r^m-r^{m'})^2}}{|\mf R|(|\mf R|^2-1)}$$
This coefficient depends only on the difference in ranks of the element (rule) in the two lists, and not on the ranks themselves.
Hence, the penalization is the same for differences occurring at the beginning or at the end of the lists.

\subsubsection{Kendall's $\tau$ rank correlation coefficient}

{\em Kendall's $\tau$ rank correlation coefficient}~\cite{kendall1938measure} is based on the idea of agreement among element (rule) pairs.
A rule pair is said to be \emph{concordant} if their order is the same in $L_{\mf R}^m$ and $L_{\mf R}^{m'}$, and \emph{discordant} otherwise.
$\tau$ computes the difference between the number of concordant and discordant pairs and divides by the total number of pairs as shown below.
$$\tau(L_{\mf R}^m,L_{\mf R}^{m'}) = \frac{|C|-|D|}{\frac{1}{2}|\mf R|(|\mf R|-1)}$$

\begin{equation*}
\begin{split}
    C=\{(r_i,r_j)|& r_i,r_j\in \mf R\wedge i<j\wedge \\
               &\sign({r_i^m-r_j^m}) = \sign(r_i^{m'}-r_j^{m'})\}
\end{split}
\end{equation*}
\begin{equation*}
\begin{split}
    D=\{(r_i,r_j)|& r_i,r_j\in \mf R\wedge i<j\wedge \\
               &\sign({r_i^m-r_j^m}) \neq \sign(r_i^{m'}-r_j^{m'})\}
\end{split}
\end{equation*}
Similar to {\em Spearman's}, $\tau$ varies between $1$ and $-1$, and penalizes uniformly across all positions.

\begin{figure*}[bth]
    \centering
	\subfloat[][\label{fig:hclust:ndcc:target}$\mathit{NDCC}$]{
  \begin{tikzpicture}[>=latex,every node/.style={draw,inner sep=0.05cm,outer sep=0},every text node part/.style={align=center}]
  	\node[ellipse,fill=black] (root) at (0,0) {\textcolor{white}{-0.27}};
  	\node[ellipse,below right = 3cm and 0.8cm of root] (n29) {0.49};
  	\draw[<-] (root) -- (n29);
  	\node[ellipse, above right = 0cm and 0.5cm of root] (n28) {0.42};
  	\draw[<-] (root) -- (n28);
  	\node[fill={rgb:gray,1;white,2},ellipse, above right =0cm and 0.5cm of n29] (n22) {0.78\\$G_5$};
  	\draw[<-] (n29) -- (n22);
  	\node[fill={rgb:gray,1;white,2},rectangle, above right= -0.4cm and 0.6cm of n22] (gini) {Piatetsky-Shapiro\\Specificity};
  	\draw[<-] (n22) -- (gini);
  	\node[fill={rgb:gray,1;white,2},rectangle, below right =  -0.2cm and 1.9cm  of n22] (n20) {Klosgen};
  	\draw[<-] (n22) -- (n20);
  	\node[fill={rgb:blue,1;white,2},rectangle, below right = 0.1cm and 3.1cm of n29] (phifisher) {$G_6$};
  	\draw[<-] (n29) -- (phifisher);
  	\node[ellipse, below right = 1.2cm and 0.9cm of n28] (n26) {0.82};
  	\draw[<-] (n28) -- (n26);
  	\node[fill={rgb:green,1;white,2},ellipse, above right = 0cm and 0.5cm of n26] (n23) {0.84\\$G_3$};
  	\draw[<-] (n26) -- (n23);
  	\node[fill={rgb:green,1;white,2},rectangle, below right = -0.17cm and 0.5cm of n23] (piatetsky) {Jaccard};
  	\draw[<-] (n23) -- (piatetsky);
  	\node[fill={rgb:green,1;white,2},rectangle, above right = -0.4cm and 0.5cm of n23] (klosgen) {$G_3$\\\sout{Jaccard}};
  	\draw[<-] (n23) -- (klosgen);
  		\node[fill={rgb:brown,1;white,2},rectangle, below right = 0.05cm and 1.74cm of n26] (n24) {$G_4$};
  	\draw[<-] (n26) -- (n24);
  	\node[ellipse,above right = 0cm and 2.2cm of n28] (n27) {0.85};
  	\draw[<-] (n28) -- (n27);
  	\node[fill={rgb:red,1;white,2},rectangle, below right = -0.1cm and 0.45cm of n27] (contrad) {$G_2$};
  	\draw[<-] (n27) -- (contrad);
  	\node[fill={rgb:yellow,1;white,2},rectangle, above right = -0.1cm and 0.45cm of n27] (n24) {$G_1$};
  	\draw[<-] (n27) -- (n24);
  \end{tikzpicture}
    }
    \hfill
    \subfloat[][\label{fig:hclust:tau:target}$\tau$]{
    \begin{tikzpicture}[>=latex,every node/.style={draw,inner sep=0.05cm,outer sep=0},every text node part/.style={align=center}]
    	\node[ellipse,fill=black] (root) at (0,0) {\textcolor{white}{-0.16}};
    	\node[fill={rgb:blue,1;white,2},ellipse,below right = 2.8cm and 2.cm of root] (n29) {0.63\\$G_6$};
    	\draw[<-] (root) -- (n29);
    	\node[fill={rgb:blue,1;white,2},rectangle, below right = -0.3cm and 3.5cm of n29] (phifisher) {Recall};
    	\draw[<-] (n29) -- (phifisher);
    	\node[fill={rgb:blue,1;white,2},rectangle, above right =-0.3cm and 3.5cm of n29] (n22) {Collective strength};
    	\draw[<-] (n29) -- (n22);
    	\node[ellipse, above right = 0.5cm and 0.5cm of root] (n28) {0.50};
    	\draw[<-] (root) -- (n28);
    	\node[ellipse, below right = 1.2cm and 1cm of n28] (n26) {0.70};
    	\draw[<-] (n28) -- (n26);
    	\node[fill={rgb:green,1;white,2},ellipse, above right = 0cm and 0.5cm of n26] (n23) {0.75\\$G_3$};
    	\draw[<-] (n26) -- (n23);
    	\node[fill={rgb:green,1;white,2},rectangle, below right = -0.3cm and 2.1cm of n23] (piatetsky) {Jaccard};
    	\draw[<-] (n23) -- (piatetsky);
    	\node[fill={rgb:green,1;white,2},rectangle, above right = -0.3cm and 2.1cm of n23] (klosgen) {$G_3$\\\sout{Jaccard}};
    	\draw[<-] (n23) -- (klosgen);
    	\node[ellipse, below right = 0.5cm and 0.8cm of n26] (n24) {0.77};
    	\draw[<-] (n26) -- (n24);
    	\node[fill={rgb:gray,1;white,2},rectangle, above right =  0.2cm and 1.8cm  of n24] (n20) {Klosgen};
    	\draw[<-] (n24) -- (n20);
    	\node[ellipse, below right = 0.2cm and 0.6cm of n24] (n9884) {0.86};
    	\draw[<-] (n24) -- (n9884);
    	\node[fill={rgb:gray,1;white,2},rectangle, above right= -0.3cm and 0.5cm of n9884] (gini) {Piatetsky-Shapiro\\Specificity};
    	\draw[<-] (n9884) -- (gini);
    	\node[fill={rgb:brown,1;white,2},rectangle, below right = 0.1cm and 0.5cm of n9884] (n9876) {$G_4$};
    	\draw[<-] (n9884) -- (n9876);
    	\node[ellipse,above right = 0.2cm and 0.5cm of n28] (n27) {0.60};
    	\draw[<-] (n28) -- (n27);
    	\node[fill={rgb:red,1;white,2},rectangle, below right = -0.1cm and 3.8cm of n27] (contrad) {$G_2$};
    	\draw[<-] (n27) -- (contrad);
    	\node[fill={rgb:yellow,1;white,2},rectangle, above right = -0.1cm and 3.8cm of n27] (n24) {$G_1$};
    	\draw[<-] (n27) -- (n24);
    \end{tikzpicture}
    }
   \caption{Hierarchical clustering of interestingness measures for a single target\label{fig:hclust:all:target}}
\end{figure*}
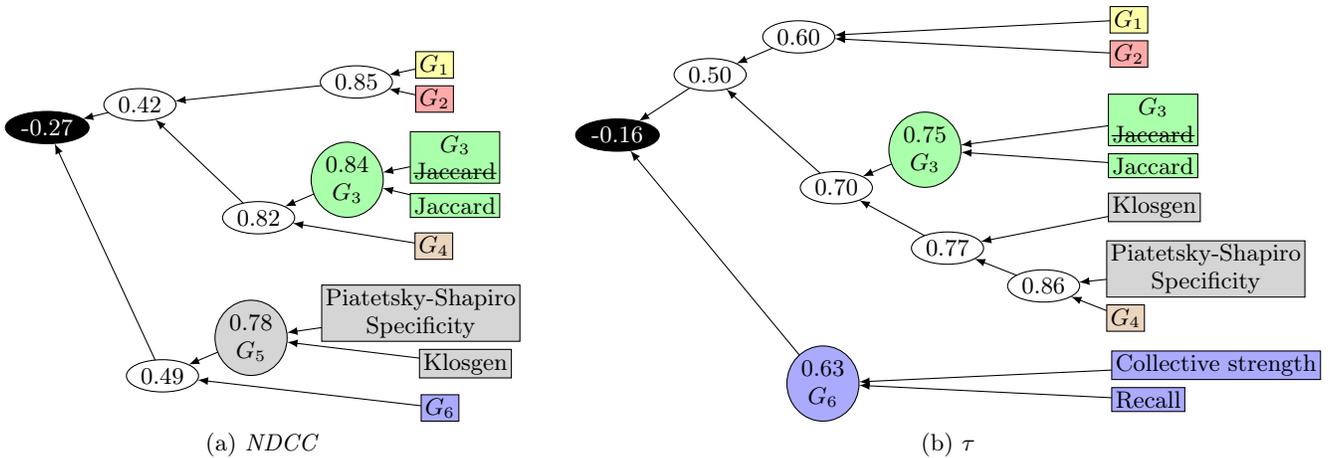

\subsubsection{Overlap@$k$}
Overlap@$k$ is another method for ranked lists comparison widely used in Information Retrieval.
It is based on the premise that in long ranked lists, the analyst is only expected to look at the top few results that are highly ranked.
While $\mathit{Spearman}$ and $\tau$ account for all elements uniformly, Overlap@$k$ compares two rankings by computing the overlap between their top-$k$ elements only.
$$\mathit{Overlap@k(L_{\mf R}^m,L_{\mf R}^{m'})=\frac{|\{r\in \mf R| r^m\leq k\}\cap\{r\in \mf R|r^{m'}\leq k\}|}{k}}$$
\subsubsection{Normalized Discounted Correlation Coefficient}
Overlap@$k$, {\em Spearman}'s and $\tau$ sit at two different extremes.
The former is conservative in that it takes into consideration only the top $k$ elements of the list
whereas the latter two take too liberal an approach by penalizing all parts of the lists uniformly.
In practice, we aim for a good tradeoff between these extremes.

To bridge this gap, we propose a new ranking correlation measure coined \emph{Normalized Discounted Correlation Coefficient} or {\em NDCC}.
{\em NDCC} draws inspiration from {\em NDCG, Normalized Discounted Cumulative Gain}~\cite{Jarvelin:2002:CGE:582415.582418}, a ranking measure commonly used in Information Retrieval.
The core idea in {\em NDCG} is to reward a ranked list $L_{\mf R}^m$ for placing an element $r$ of relevance $\mathit{rel_r}$ by $\frac{\mathit{rel_r}}{\log{r^m}}$.

The logarithmic part acts as a smoothing discount rate representing the fact that as the rank increases, the analyst is less likely to observe $r$.
In our setting, there is no ground truth to properly assess $\mathit{rel_r}$.
Instead, we use the ranking assigned by $m'$ as a relevance measure for $r$, with an identical logarithmic discount.
When summing over all of $\mf R$, we obtain $\mathit{DCC}$, which presents the advantage of being a symmetric correlation measure between two rankings $L_{\mf R}^m$ and $L_{\mf R}^{m'}$.
$$\mathit{DCC(L_{\mf R}^m,L_{\mf R}^{m'})}=\sum_{r\in \mf R}{\frac{1}{\log{(1+r^{m'})}\log{(1+r^m)}}}$$
We compute $\mathit{NDCC}$ by normalizing $\mathit{DCC}$ between $1$ (identical rankings) and $-1$ (reversed rankings).
$$\mathit{NDCC(L_{\mf R}^m,L_{\mf R}^{m'})}=\frac{dcc-avg}{max-avg}$$
\begin{equation*}
\begin{split}
\mbox{where } dcc&=\mathit{DCC(L_{\mf R}^m,L_{\mf R}^{m'})},\ max=\mathit{DCC(L_{\mf R}^{m'},L_{\mf R}^{m'})}\\
    min&=\mathit{DCC(L*,L_{\mf R}^{m'})},\ L*=\mathit{rev(L_{\mf R}^{m'})}\\
    avg&=(max+min)/2
\end{split}
\end{equation*}
\subsubsection{Ranking comparison by example}
We illustrate the difference between all ranking correlation measures with an example in Table~\ref{tab:exampleranks}.
This shows correlation of a ranking $L^1$ with 3 others, according to each measure.
$\mathit{NDCC}$ does indeed penalize differences at higher ranks, and is more tolerant at lower ranks.

\subsection{Ranking rules with identical targets}
\label{sec:identicaltargets}
We first consider the case of ranking association rules $A\rightarrow B$ where $B$ is a product, i.e., all rules have the same $B$.
We perform a comparative analysis of ranking measures on our 3 mining scenarios summarized in Table~\ref{tab:scenarios}.
Our first observation is that the results we obtain for all scenarios lead to the same conclusions. Therefore, we only report numbers for \prodassocclient.
We use as targets for this comparison 64 products previously studied by analysts that lead to the discovery of \num{1651024} association rules.
We compute one rule ranking per interestingness measure.

While all measures are computed differently, we notice that some of them always return the  same ranking for association rules of a given target.
We identify them in Table~\ref{tab:measures} using symbols.
Other notable similarities include {\em Sebag-Schoenauer} and {\em lift} (89\% of rankings are equal), as well as {\em Loevinger} and {\em lift} (87\%).
This difference between the number of interestingness measures considered (34) and the number of different rankings obtained (25) can easily be explained analytically in the case of a fixed target. Indeed, for a given ranking, $P(B)$ is constant, which eliminates some of the differences between interestingness measures.
In addition, some measures only have subtle differences which only appear when selecting extreme values for $P(A)$, $P(B)$ and $P(AB)$, which do not occur in practice in our retail dataset.

\begin{table}[t]
\begin{center}
\begin{tabular}{c|c}
Ranking & Content\\
\hline
$L^1$& $r_1,r_2,r_3,r_4$ \\
$L^2$& $r_2,r_1,r_3,r_4$ \\
$L^3$& $r_1,r_2,r_4,r_3$  \\
$L^4$& $r_2,r_3,r_1,r_4$ \\
\end{tabular}
\begin{tabular}{c|c|c|c|c}
&$\mathit{Spearman}$&$\tau$&$\mathit{Overlap}$@$2$&$\mathit{NDCC}$\\
\hline
$L^2$&$0.80$&$0.67$&$1$&$0.20$\\
$L^3$&$0.80$&$0.67$&$1$&$0.97$\\
$L^4$&$0.40$&$0.33$&$0.5$&$-0.18$\\
\end{tabular}
\caption{Example rankings and correlations\label{tab:exampleranks}}
\end{center}
\end{table}

\subsubsection{Comparative analysis}
We now evaluate similarity between interestingness measures that do not return the same rankings.
We compute a $34\times34$ correlation matrix of all rankings according to each correlation measure described in Section~\ref{sec:correlations}, and average them over the 64 target products.
This gives us a ranking similarity between all pairs of measures
We then rely on hierarchical clustering with average linkage~\cite{sokal58} to obtain a dendrogram of interestingness measures and analyze their similarities.
The dendrograms for $\mathit{NDCC}$ and $\tau$ are presented in Figure~\ref{fig:hclust:all:target}.
For better readability, we merge sub-trees when correlation is above $0.9$.
To describe the results more easily, we partition interestingness measures into 6 groups, as indicated in the third column in Table~\ref{tab:measures}.

\begin{figure}[t]
\begin{center}
\includegraphics[scale=1.2]{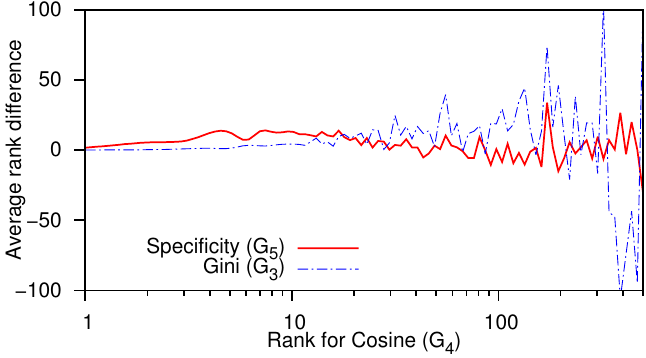}
\caption{Rank correlations\label{fig:rankcorrelation}}
\end{center}
\end{figure}

$G_1$ is by far the largest group: in addition to 4 measures that always generate the same rankings, 14 other measures output similar results.
A second group, $G_2$, comprising 2 measures, is quite similar to $G_1$ according to $\mathit{NDCC}$.
$\tau$ also discovers this similarity, but considers it lower, which shows that it is mostly caused by high ranks.
{\em Jaccard} is as a slight outlier in $G_3$ according to $\mathit{NDCC}$.
Indeed, when focusing on the first 20 elements ($\mathit{Overlap}$@$20$), only an average of 71\% are shared between {\em Jaccard} and the rest of $G_3$.
This situation also occurs between {\em Klosgen} and the rest of $G_5$.
Interestingly, we observe that, according to $\mathit{NDCC}$, $G_5$ is closest to $G_6$ and is negatively correlated with the other groups.
However, according to $\tau$, $G_5$ is very similar to $G_4$ and is negatively correlated with $G_6$.
This difference of behavior between ranking measures illustrates the importance of accounting for rank positions.
When the top of the ranking is considered more important, some similarities emerge.
We illustrate this behavior in Figure~\ref{fig:rankcorrelation} by displaying correlation between rankings obtained with different interestingness measures.
This experiment clearly shows that overall, {\em cosine} ($G_4$) is closer to {\em specificity} ($G_5$) than {\em Gini} ($G_3$), as the rank difference observed in the results is overall smaller.
However, when focusing on the top-10 results of {\em cosine}, {\em Gini} assigns closer ranks than {\em specificity}.
This explains the difference in clustering between $\mathit{NDCC}$/$\mathit{overlap}$ and $\tau$/$\mathit{Spearman}$.

\subsubsection{Annotating groups}
\label{sec:annotation}
\begin{figure}[t]
\centering
\includegraphics[scale=1.12]{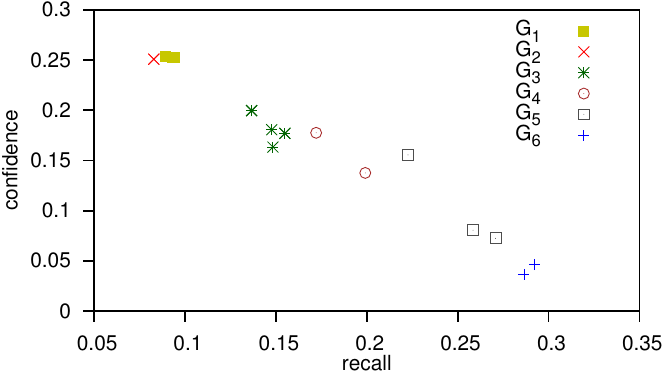}
\caption{Average recall/confidence of the top-20 results of interestingness measures\label{fig:recallprecision}}
\end{figure}

While using hierarchical clustering on interestingness measures allows the discovery of families of measures, and their relative similarity, it does not fully explain which types of results are favored by each of them.
We propose to compare their outputs according to the two most basic and intuitive interestingness measures employed in data mining: {\em recall} and {\em confidence}.
{\em recall} represents the proportion of target items that can be retrieved by a rule, that is,  $P(A|B)$.
Its counterpart, {\em confidence}, represents how often the consequent is present when the antecedent is, that is, $P(B|A)$.
We present, in Figure~\ref{fig:recallprecision}, the average {\em recall} and {\em confidence} of the top-20 rules ranked according to each interestingness measure.
$G_1$ contains {\em confidence}, so it is expected to score the highest on this dimension.
$G_2$ is extremely close to $G_1$, but obtains slightly lower {\em confidence} and {\em recall}.
We then have, in order of increasing {\em recall} and decreasing {\em confidence} $G_3$, $G_4$ and $G_5$.
Finally, $G_6$, which contains {\em recall}, obtains the highest {\em recall} but the lowest {\em confidence}.
Figure~\ref{fig:recallprecision} also shows that executing a Euclidean distance-based clustering, such as $k$-means, with recall/confidence coordinates would lead to groups similar to the ones obtained with hierarchical clustering.
Hence, this analysis is consistent with the hierarchical grouping and the correlation with $\mathit{NDCC}$.


While we believe that $\mathit{NDCC}$ reflects better the interpretation of analysts browsing rules,
it is important to note that the grouping of interestingness measures created through this evaluation
is stable across all 4 correlation measures and for all 3 scenarios.
Correlation between different families of measures may vary, but measures within a single family always have a high similarity.
Thus, we can safely state that the obtained results are true in the general case of food retailers and we can rely on these groups to reduce the number of options presented to analysts.

%

\subsection{Ranking rules with  different targets}
\label{sec:differenttargets}
\begin{figure*}[th!]
    \centering
	\subfloat[][\label{fig:hclust:ndcc:global}$\mathit{NDCC}$]{

  \begin{tikzpicture}[>=latex,every node/.style={draw,inner sep=0.05cm,outer sep=0},every text node part/.style={align=center}]

      \node[ellipse,fill=black] (root) at (0,0) {\textcolor{white}{-0.45}};


          \node[ellipse,xshift=0.5cm,yshift=2cm] (n4) at (root) {0.27};
          \draw[<-] (root) -- (n4);

              \node[ellipse,xshift=1.5cm,yshift=2cm] (n6) at (n4) {0.65};
              \draw[<-] (n4) -- (n6);

                  \node[ellipse, xshift=1.5cm,yshift=-0.6cm] (n7) at (n6) {0.82};
                  \draw[<-] (n6) -- (n7);

                      \node[rectangle, xshift=2.6cm,yshift=-0.2cm,fill=g2Color] (lsct) at (n7) {Least Contradiction};
                      \draw[<-] (n7) -- (lsct);

                      \node[rectangle, xshift=1.3cm,yshift=0.3cm,fill=g1aColor] (mix1) at (n7) {$G_1^a$};
                      \draw[<-] (n7) -- (mix1);

                  \node[rectangle, xshift=2.8cm,yshift=0.4cm,fill=g1bColor] (mix2) at (n6) {$G_1^b$};
                  \draw[<-] (n6) -- (mix2);

              \node[ellipse,xshift=1 cm,yshift=-1.1cm] (n5) at (n4) {0.53};
              \draw[<-] (n4) -- (n5);

                  \node[ellipse,xshift=1.2 cm,yshift=0.4cm] (n9) at (n5) {0.74};
                  \draw[<-] (n5) -- (n9);

                  \node[ellipse,xshift=1.7 cm,yshift=-1.5cm] (n8) at (n5) {0.78};
                  \draw[<-] (n5) -- (n8);

                      \node[rectangle, xshift=2.0cm,yshift=-0.2cm,fill=g3Color] (jaccard) at (n8) {Jaccard};
                      \draw[<-] (n8) -- (jaccard);

                      \node[rectangle, xshift=2.8cm,yshift=0.3cm,fill=g5Color] (pshp) at (n8) {Piatetsky-Shapiro};
                      \draw[<-] (n8) -- (pshp);

                      \node[ellipse,xshift=1 cm,yshift=0.5cm] (n10) at (n9) {0.87};
                      \draw[<-] (n9) -- (n10);

                          \node[rectangle, xshift=1.1cm,yshift=-0.5cm,fill=g4Color] (mix4) at (n10) {$G_4$};
                          \draw[<-] (n10) -- (mix4);

                          \node[rectangle, xshift=1.5cm,yshift=0.4cm,fill=g3Color] (mix5) at (n10) {$G_3$\\\sout{Jaccard}\\\sout{Gini}};
                          \draw[<-] (n10) -- (mix5);

                      \node[rectangle, xshift=2.5cm,yshift=-0.7cm] (mix3) at (n9) {Klosgen\\Gini};
                      \draw[<-] (n9) -- (mix3);

          \node[ellipse,xshift=1cm,yshift=-1.8cm] (n1) at (root) {-0.01};
          \draw[<-] (root) -- (n1);

              \node[rectangle, xshift=4.4cm,yshift=0.35cm,fill=g2Color] (acc) at (n1) {Accuracy};
              \draw[<-] (n1) -- (acc);

              \node[ellipse, xshift=1.6cm,yshift=-0.5cm] (n2) at (n1) {0.73};
              \draw[<-] (n1) -- (n2);

                  \node[rectangle, xshift=2.85cm,yshift=0.25cm,fill=g5Color] (specif) at (n2) {Specificity};
                  \draw[<-] (n2) -- (specif);

                  \node[rectangle, xshift=2.25cm,yshift=-0.25cm, fill=g6Color] (n3) at (n2) {$G_6$};
                  \draw[<-] (n2) -- (n3);
  \end{tikzpicture}

    }
    \hfill
    \subfloat[][\label{fig:hclust:tau:global}$\tau$]{

    \begin{tikzpicture}[>=latex,every node/.style={draw,inner sep=0.05cm,outer sep=0},every text node part/.style={align=center}]

        \node[ellipse,fill=black] (root) at (0,0) {\textcolor{white}{-0.14}};


            \node[ellipse,xshift=0.7cm,yshift=2.2cm] (n4) at (root) {0.47};
            \draw[<-] (root) -- (n4);

                \node[ellipse,xshift=1.2cm,yshift=2cm] (n6) at (n4) {0.59};
                \draw[<-] (n4) -- (n6);

                    \node[ellipse, xshift=1cm,yshift=-1cm] (n7) at (n6) {0.69};
                    \draw[<-] (n6) -- (n7);

                        \node[ellipse, xshift=1.2cm,yshift=0.4cm] (n5757) at (n7) {0.75};
                        \draw[<-] (n7) -- (n5757);
                        \node[rectangle, xshift=3.6cm,yshift=-1.cm,fill=g2Color] (lsct) at (n7) {Least contradiction};
                        \draw[<-] (n7) -- (lsct);

                       	 \node[rectangle, xshift=2.5cm,yshift=0.3cm,fill=g1aColor] (mix1) at (n5757) {$G_1^a$};
                     	 \draw[<-] (n5757) -- (mix1);
                       	 \node[ellipse, xshift=1.2cm,yshift=-0.6cm] (n8754) at (n5757) {0.87};
    			 \draw[<-] (n5757) -- (n8754);
    				 \node[rectangle, xshift=1.4cm,yshift=-0.25cm,fill=g3Color] (gini) at (n8754) {Gini};
    				 \draw[<-] (n8754) -- (gini);
    				 \node[rectangle, xshift=1.7cm,yshift=0.25cm,fill=g5Color] (klosgen) at (n8754) {Klosgen};
    				 \draw[<-] (n8754) -- (klosgen);
                    \node[rectangle, xshift=4.7cm,yshift=0.5cm,fill=g1bColor] (mix2) at (n6) {$G_1^b$};
                    \draw[<-] (n6) -- (mix2);

                \node[ellipse,xshift=2 cm,yshift=-2cm] (n5) at (n4) {0.67};
                \draw[<-] (n4) -- (n5);

                    \node[ellipse,xshift=2.2 cm,yshift=0.4cm] (n9) at (n5) {0.82};
                    \draw[<-] (n5) -- (n9);
    			\node[rectangle, xshift=1.7cm,yshift=-0.35cm,fill=g4Color] (mix4) at (n9) {$G_4$};
                             \draw[<-] (n9) -- (mix4);
    			\node[rectangle, xshift=2.1cm,yshift=0.55cm,fill=g3Color] (mix3) at (n9) {$G_3$\\\sout{Jaccard}\\\sout{Gini}};
                             \draw[<-] (n9) -- (mix3);

                   \node[rectangle, xshift=4cm,yshift=-0.5cm,fill=g5Color] (pshp) at (n5) {Piatetsky-Shapiro};
                   \draw[<-] (n5) -- (pshp);


    %
    %
    %
    %
    %
    %

            \node[ellipse,xshift=0.5cm,yshift=-1.4cm] (n1) at (root) {0.35};
            \draw[<-] (root) -- (n1);

                \node[rectangle, xshift=6.6cm,yshift=0.35cm,fill=g2Color] (acc) at (n1) {Accuracy};
                \draw[<-] (n1) -- (acc);
    	    \node[ellipse,xshift=1.3cm,yshift=-0.5cm] (n498156) at (n1) {0.62};
                \draw[<-] (n1) -- (n498156);
                    \node[rectangle, xshift=5.35cm,yshift=0.25cm,fill=g5Color] (specif) at (n498156) {Specificity};
                    \draw[<-] (n498156) -- (specif);
                \node[ellipse, xshift=1.0cm,yshift=-0.5cm] (n2) at (n498156) {0.68};
                \draw[<-] (n498156) -- (n2);
                    \node[rectangle, xshift=4.15cm,yshift=0.2cm,fill=g3Color] (jaccard) at (n2) {Jaccard};
                     \draw[<-] (n2) -- (jaccard);

                    \node[rectangle, xshift=3.8cm,yshift=-0.25cm, fill=g6Color] (n3) at (n2) {$G_6$};
                    \draw[<-] (n2) -- (n3);
    \end{tikzpicture}

    }
   \caption{Hierarchical clustering of interestingness measures for multiple targets\label{fig:hclust:global}}
\end{figure*}
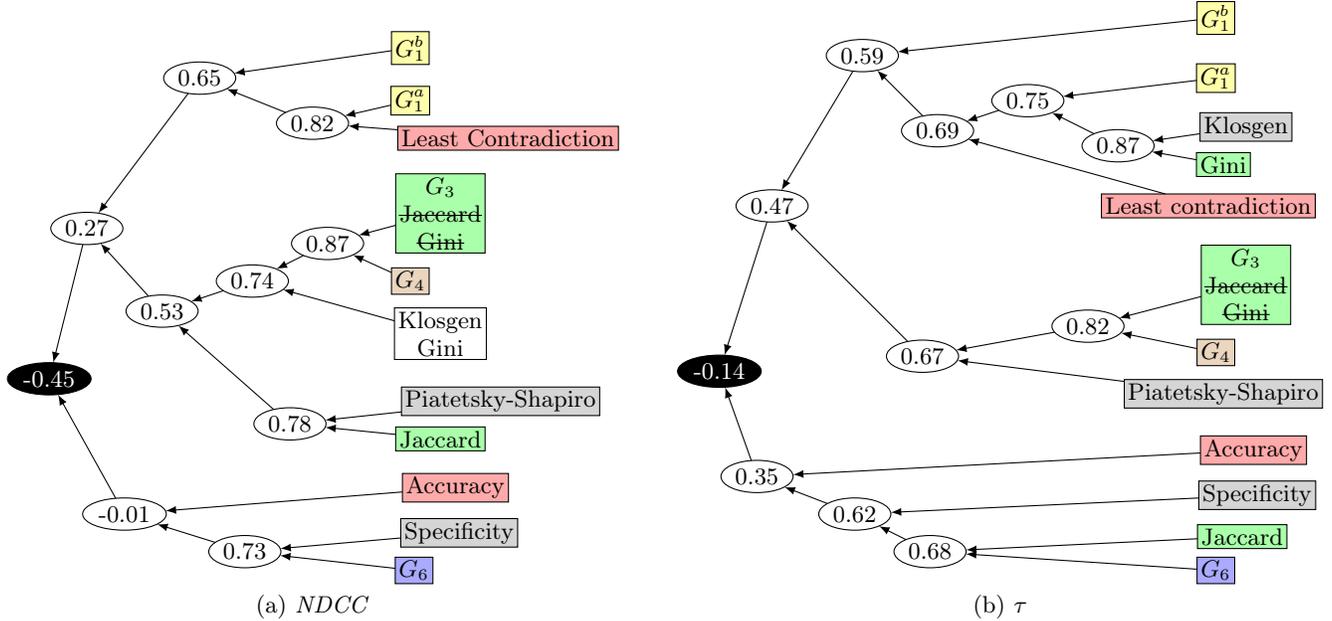

We now consider the problem of ranking association rules when many targets are provided as input, i.e. association rules $A \rightarrow B$ can have different targets $B$.
Compared to having a single target, this setting introduces one more degree of freedom in the quality measure of the association rules, as $P(B)$ varies.
We rely on the same set of (64) products as in the identical target experiment, but instead of generating rankings for each target, we rank all association rules together.
The dendrogram of quality measures obtained for $\mathit{NDCC}$ and $\tau$ is presented in Figure~\ref{fig:hclust:global}.

We observe a much wider variety in rankings. 
The group $G_1$, observed previously, splits into two different sub-groups, $G_1^a$ and $G_1^b$.
A large fraction of $G_3$ remains similar, and the two measures that constitute $G_4$ and $G_6$ remain highly correlated.
$G_2$ and $G_5$ are not preserved when ranking different targets simultaneously.

We observe a stronger agreement between $\mathit{NDCC}$ and $\tau$.
The only notable differences are \textit{(i)}  {\em Klosgen} and {\em Gini}, which are highly correlated with $G_4$ when focusing on the top results, while they are more similar to $G_1^a$ globally, and \textit{(ii) Jaccard}, which switches from a similarity with {\em Piatetsky-Shapiro} to a similarity with $G_6$.

Measures that prioritize high values of $P(B)$, i.e. favor targets that are more frequent, are $G_1^a$, {\em Piatetsky-Shapiro}, {\em Klosgen} and {\em Gini}.
Indeed, in the case of {\em confidence} ($G_1^a$), an association rule $A\rightarrow B$ that has a very frequent $B$ can easily score highly by selecting a very specific $A$.
Conversely, {\em specificity}, {\em collective strength}, {\em accuracy}, $G_1^b$ and {\em recall} tend to rank less frequent targets highly.
A similar explanation applies to {\em recall}, as a low frequency of $B$ makes it easier to find association rules that capture most of its appearances in the data.

\subsection{Selecting representative measures}
\label{sec:selection}
We summarize the findings of the comparative evaluation in Table~\ref{tab:groupsummary}.
When the analyst selects a single target, we identify 6 families of measures that behave similarly.
When multiple targets are selected, $G_1$ splits into 2 sub-groups.
Each family offers a different trade-off in terms of confidence and recall, and thus ranks association rules differently.
When ranking rules with different targets, some families are sensitive to the frequency of the target.

We select the quality measure that most represents each family of measures (i.e. with highest average similarity) in order to confront the results of this analysis with the opinion of domain experts in our user study.
Taking a general data mining perspective leads us to considering $G_3$ and $G_4$ as the most promising families for finding interesting association rules.
Indeed, it is important to achieve a good trade-off between {\em recall} and {\em confidence} in order to find reliable association rules that can be applied in a significant number of cases.
Hence, {\em F1} score, that combines {\em recall} and {\em confidence}, would prefer $G_3$ and $G_4$ to others.

\begin{table}[bt]
\centering
\begin{tabular}{|l|l|}
\hline
\begin{tabular}{l}Group\\Representative\end{tabular}&Description\\\hline
\cellcolor{g1aColor}\begin{tabular}{l}$G_1^a$\\Lift\end{tabular}&\begin{tabular}{l}Highest confidence\\Very low recall\\Favors frequent targets\end{tabular}\\\hline
\cellcolor{g1bColor}\begin{tabular}{l}$G_1^b$\\Added value\end{tabular}&\begin{tabular}{l}Highest confidence\\Very low recall\\Favors rare targets\end{tabular}\\\hline
\cellcolor{g2Color}\begin{tabular}{l}$G_2$\\Accuracy\end{tabular}&\begin{tabular}{l}Very high confidence\\Very low recall\end{tabular}\\\hline
\cellcolor{g3Color}\begin{tabular}{l}$G_3$\\Fisher's exact test\end{tabular}&\begin{tabular}{l}High confidence\\Low recall\\Low sensitivity to target freq.\end{tabular}\\\hline
\cellcolor{g4Color}\begin{tabular}{l}$G_4$\\Cosine\end{tabular}&\begin{tabular}{l}Average confidence\\Average recall\\Low sensitivity to target freq.\end{tabular}\\\hline
\cellcolor{g5Color}\begin{tabular}{l}$G_5$\\Piatetsky-Shapiro\end{tabular}&\begin{tabular}{l}Low confidence\\High recall\end{tabular}\\\hline
\cellcolor{g6Color}\begin{tabular}{l}$G_6$\\Collective strength\end{tabular}&\begin{tabular}{l}Lowest confidence\\Highest recall\\Favors rare targets\end{tabular}\\\hline
\end{tabular}
\caption{Summary of quality measure groups\label{tab:groupsummary}}
\end{table}

\section{User study}
\label{sec:xp:user}
We now report the results of a user study with domain experts from Intermarch\'e.
The goal of this study is to assess the ability of interestingness measures to rank association rules according to the needs of an analyst.
As explained in Section~\ref{sec:xp:empirical}, we identified 6 families of measures, and selected a representative of each group for the user study (Table~\ref{tab:groupsummary}).
We rely on the expertise of our industrial partner to determine, for each analysis scenario, which family produces the most interesting results.
This experiment involved 2 experienced analysts from the marketing department of Intermarch\'e.
We setup \project and let analysts select targets multiple times in order to populate the web application's database with association rules (Section~\ref{sec:exploitation}).
We let our analysts interact with \project without any time restriction, and collect their feedback in a free text form.

Each analyst firstly has to pick a mining scenario among \demoassoc, \prodassocreceipt, or \prodassocclient.
Then she picks a target category or a target product in the taxonomy.
In \prodassocreceipt\ and \prodassocclient, she also has the option to
filter out rules whose antecedent products are not from the same category as the target.
Finally, she chooses one of our 6 ranking measures to sort association rules.
Neither the name of the measure nor its computed values for association rules are revealed,
because we wanted analysts to evaluate rankings without knowing how they were produced.

Resulting association rules are ranked according to a selected measure.
Each rule is displayed with its support, confidence and recall,
such that analysts can evaluate it at a glance.
For each scenario,
our analysts are asked which representative measure highlights the most interesting results
(as detailed below, in all cases a few of them were chosen).

\subsection{Scrolling behavior}
Once the analyst selects a target, \emph{all} matching rules are returned.
The initial motivation of this choice was to determine how many results
are worth displaying and are actually examined by the analysts.
According to the follow-up interview with the analysts, they carefully considered the first ten results,
and screened up to a hundred more.
Interestingly, analysts mentioned that
they also scrolled down to the bottom of the list
in order to see which customer segments are not akin to buying the selected category.
For example, when browsing demographic association rules, they expected to find \{{\em 50-64}\} $\rightarrow$ {\em pet food} among top results,
but also expected \{{\em <35, Paris}\} $\rightarrow$ {\em pet food} among bottom results.
This confirms that all rules should remain accessible.
This also indicates that while interestingness measures favor strong associations, it would also be interesting to highlight {\em anti}-rules, as those can also convey useful information.

\subsection{Feedback on ranking measures}
We let marketing experts explore all 3 scenarios and express their preference towards groups of measures.

In the \demoassoc case, $G_1$ and $G_3$ were both highly appreciated.
$G_1$ favors rules such as $\{< 35, M, $ Oise$\}\rightarrow$ {\em Flat and Carbonated drinks}.
These rules are very specific and thus have a very high confidence (31,58 \% in this particular case).
However, this comes at the cost of recall (0,08 \%).
Experts value {\em confidence} much more than {\em recall}, as their priority is finding rules that they consider reliable.
A low support is not necessarily an issue, and can lead to the discovery of surprising niche rules that can be exploited nonetheless.
As discussed in Section~\ref{sec:annotation}, $G_3$ offers a more balanced trade-off between confidence and recall, and prioritizes rules such as $\{$< 35, *, *$\}\rightarrow$ {\em Baby food} (confidence 8,57 \%, recall 37,61\%).
These rules are interesting because they capture a large fraction of the sales of a given category, but are less reliable and generally less surprising.
$G_2$ and $G_4$ were considered as less interesting than $G_1$ and $G_3$ respectively.
Their results offer similar trade-offs, but with lower confidence each time.
$G_5$ and $G_6$ were considered unusable because of their very low confidence.

When experimenting with \prodassocall, we observed a slightly different behavior.
By default, the analysts favored $G_1$ and $G_2$ because of the confidence of their results.
Then, we offered the analysts the possibility of filtering the rules to only keep the ones in which the antecedent contains products from the same category as the target.
This led to analysts favoring $G_3$ and $G_5$.
This difference is caused by an important but implicit criterion: the ability of a measure to filter out very popular products.
For example, the rule \{{\em vanilla cream, emmental}\}$\rightarrow$ {\em chocolate cream}
usually appears just above its shorter version
\{{\em vanilla cream}\}$\rightarrow$ {\em chocolate cream},
because the first one has a confidence of $32\%$ and the second $31\%$.
However, experts prefer the second one, because \textit{emmental} (cheese)
is among the heavy hitters in stores. Its addition to the rule is hence considered insignificant.
This ``noise'' generally increases with {\em recall}.
Hence, when no filtering is available, $G_1$ is selected, but analysts prefer the {\em recall} and {\em confidence} trade-off provided by $G_3$ and $G_5$.
Again, $G_4$ suffered from its proximity to $G_3$ with lower confidence, while $G_6$'s confidence was too low.

\subsection{Summary of user study}
In all cases, analysts mentioned $G_6$
as uninteresting overall because it selects rules of low {\em confidence}.
In general, sorting by decreasing {\em lift}
(which is close to sorting by decreasing {\em confidence})
is the preferred choice.
Combined with the minimum support threshold used in the mining phase, this ranking promotes rules that are considered reliable.
However, the preference of the analysts changes when filters are available to narrow down the set of rules to specific product categories.
In this case, they favor the compromise between {\em confidence} and {\em support} offered, for instance, by the {\em Piatetsky-Shapiro}'s measure~\cite{Piatetsky1991KDD}.

\section{Related work}
\label{sec:related}

To the best of our knowledge, \project targets datasets which are
orders of magnitude bigger (and sparser) than those tested in existing work on ranking association rules.
This paper is also the first to complement an algorithmic comparative analysis with a user study involving domain experts.

The definition of quality of association rules is a well-studied topic
in  statistics and data mining, summarized in~\cite{geng2006ACM}.
In this survey, Geng \textit{et al.} review as many as 38 measures
for association and classification rules.
They also discuss 4 sets of properties like symmetry or monotony,
and how each of them highlights different meanings of ``rule quality'', such as novelty and generality.
However, we observe no correlation between these properties and the groups of measures discovered using \project.

These 38 measures are compared in~\cite{LeSANER15}.
Authors consider the case of extracting and ranking temporal rules (\textit{event A}$\rightarrow$\textit{event B}) from the execution traces of Java programs.
Each measure is evaluated in its ability to rank highly rules known from a ground truth (Java library specification).
We observe that the measures scoring the highest are all from the groups identified in this work as $G_1$ and $G_3$, which were also favored by our analysts.
There are however some counterexamples, with measures from $G_1$ scoring poorly.
The authors then use a statistical approach to build a partial ordering of measures quality.
This results in the formation of measure equivalence classes.
However, the semantic of these classes is based on the principle of dominance in the evaluation, and not on the comparison of the rankings themselves.
Hence, the equivalence classes obtained do not match our groups.
The main difference between \project and ~\cite{LeSANER15}  is the absence of a ground truth of interesting rules for our dataset.
Consequently, our evaluation of measures is first comparative, with 4 correlations measures covering both the top of the ranking and the entire ranked list.
We then build groups of measures to reduce the number of options presented to expert analysts in the user study.
The differences in the results obtained also highlight the importance of performing domain-specific studies, as the properties of data and the expectations of analysts vary significantly.

The closest work to ours is \textsc{Herbs}~\cite{Lenca2007}.
\textsc{Herbs} relies on a different and smaller set of measures to cluster rule rankings.
Authors perform an analysis of the properties of measures, in addition to an experimental study.
The datasets used are from the health and astronomy domains.
Each of them contains at most \num{1728} transactions and leads to the extraction of 49 to \num{6312} rules.
Rankings are then compared between all pairs of measures using Kendall's $\tau$ correlation measure averaged over all datasets.
The largest group of measures identified, which includes confidence, is quite similar to $G_1$.
However, there are also significant differences.
For instance, we find $G_2$ and $G_6$ to be very different, while~\cite{Lenca2007} considers the measures of this group similar.
The authors observe a weak resemblance between the theoretical and experimental analysis of the measures.
The main similarity between~\cite{Lenca2007} and \project is the reliance on a pairwise correlation measure followed by a hierarchical clustering to detect groups of measures.
\project is entirely focused on retail data, which has different properties and contains millions of transactions and rules.
\project is also more exhaustive in the analysis of measures: we consider more interestingness measures, and 4 different ranking correlation measures instead of 1.
This allows us to discover more subtle differences in a more specific domain.
Finally, we perform a user study to assess the quality of each group according to experts from the retail industry.

Our use of the {\em p-value}
(via {\em Pearson's $\chi^2$ test}) in the evaluation of rule interestingness is borrowed from~\cite{LiuICDE2011}.
A low {\em p-value} shows a correlation between a rule's antecedent and consequent.
The use of {\em Fisher}'s exact test on association rules is inspired by~\cite{Minato2014KDD}.
Both of these works aim at finding highly-correlated itemsets,
which requires the analyst to set a threshold on the $p$-value.
This is common practice in biology,
but less meaningful in the retail industry.

In~\cite{LiuICDE2011}, Liu \textit{et al.} also propose an exploration framework
where rules are grouped by consequent, then
traversed by progressively adding items to the antecedent.
The framework provides hints to help guess how each
additional item would make a difference.
Such a framework is suitable to some of the scenarios we consider
and could be integrated in a future version of~\project.

\section{Summary and evolutions}
\label{sec:evolutions}
In this paper, we present \project, a framework for mining association rules from large-scale retail data.
We defined 3 mining scenarios allowing analysts to extract associations between user segments and product categories, or products themselves.
Given a scenario, \project builds a dataset of transactions and mines in parallel association rules containing targets selected by the analysts.
Our main contribution is the study of 34 interestingness measures for association rules.
We first performed an analytical and an empirical comparison between different rule rankings and grouped measures into 6 families.
Resulting groups were then evaluated in a user study involving retail experts.
We concluded that {\em lift} and {\em Piatetsky-Shapiro} best fit the needs of the analysts, as they ensure a high confidence.

We foresee 3 directions of improvement for \project.
The first one is related to the architecture.
\project is currently implemented using batch processing and on-disk storage.
While mining is already fast, I/O operations introduces some latency between the definition of a mining scenario and the display of results.
We are currently migrating to an in-memory dataset representation, using Spark~\cite{spark}, to allow faster target selection and lower response time.
A second improvement is the extraction of negative results ({\em anti}-rules). That is particularly true for rules containing customer segments.
We hence need to determine how negative rules should be ranked in order to properly adjust their proportion in the outcome.
Finally, while quality measures are crucial to select the most interesting results for the analysts,
we  would like to introduce diversity in displaying rules and study its impact on analysts' satisfaction.


\begin{thebibliography}{10}

\bibitem{AggrawalSIGMOD1993}
R.~Agrawal, T.~Imieli\'{n}ski, and A.~Swami.
\newblock Mining {A}ssociation {R}ules between {S}ets of {I}tems in {L}arge
  {D}atabases.
\newblock In {\em Proc. SIGMOD}, pages 207--216, 1993.

\bibitem{daniel1978applied}
W.~Daniel.
\newblock {\em Applied {N}onparametric {S}tatistics}.
\newblock Houghton Mifflin, 1978.

\bibitem{mapreduce}
J.~Dean and S.~Ghemawat.
\newblock Mapreduce: {S}implified {D}ata {P}rocessing on {L}arge {C}lusters.
\newblock {\em Commun. ACM}, 51(1):107--113, 2008.

\bibitem{SuganthanSIGMOD2015}
P. Suganthan G.C., C. Sun, K. Gayatri K., H. Zhang, F. Yang, N. Rampalli, S. Prasad, E. Arcaute, G. Krishnan, R. Deep, V. Raghavendra, A. Doan
\newblock Why {B}ig {D}ata {I}ndustrial {S}ystems {N}eed {R}ules and {W}hat
  {W}e {C}an {D}o {A}bout {I}t.
\newblock In {\em Proc. SIGMOD}, pages 265--276, 2015.

\bibitem{geng2006ACM}
L.~Geng and H.~J. Hamilton.
\newblock {I}nterestingness {M}easures for {D}ata {M}ining: {A} {S}urvey.
\newblock {\em ACM Comput. Surv.}, 38(3), 2006.

\bibitem{Minato2014KDD}
S.i. Minato, T.~Uno, K.~Tsuda, A.~Terada, and J.~Sese.
\newblock A {F}ast {M}ethod of {S}tatistical {A}ssessment for {C}ombinatorial
  {H}ypotheses based on {F}requent {I}temset {E}numeration.
\newblock {\em Lect. Notes Artif. Int.}, 8725:422--436, 2014.

\bibitem{Jarvelin:2002:CGE:582415.582418}
K.~J\"{a}rvelin and J.~Kek\"{a}l\"{a}inen.
\newblock Cumulated {G}ain-based {E}valuation of {IR} {T}echniques.
\newblock {\em ACM Trans. Inf. Syst.}, 20(4):422--446, 2002.

\bibitem{kendall1938measure}
M.~G. Kendall.
\newblock A {N}ew {M}easure of {R}ank {C}orrelation.
\newblock {\em Biometrika}, 30(1/2):81--93, 1938.

\bibitem{jlcm}
M.~Kirchgessner, V.~Leroy, A.~Termier, S.~Amer-Yahia, and M.-C. Rousset.
\newblock {jLCM}.
\newblock \url{http://slide-lig.github.io/jlcm/}.
\newblock [Online; accessed 29-Feb-2016].

\bibitem{LeSANER15}
T.-D. Le and D.~Lo.
\newblock Beyond {S}upport and {C}onfidence: {E}xploring {I}nterestingness
  {M}easures for {R}ule-{B}ased {S}pecification {M}ining.
\newblock In {\em Proc. SANER}, pages 331--340, 2015.

\bibitem{Lenca2007}
P.~Lenca, B.~Vaillant, P.~Meyer, and S.~Lallich.
\newblock Association {R}ule {I}nterestingness {M}easures: {E}xperimental and
  {T}heoretical {S}tudies.
\newblock In {\em Quality Measures in Data Mining}, pages 51--76. Springer,
  2007.

\bibitem{LiuICDE2011}
G.~Liu, M.~Feng, Y.~Wang, L.~Wong, S.-K. Ng, T.~L. Mah, and E.~J.~D. Lee.
\newblock Towards {E}xploratory {H}ypothesis {T}esting and {A}nalysis.
\newblock In {\em Proc. ICDE}, pages 745--756, 2011.

\bibitem{dice15}
S.~Mishra, V.~Leroy, and S.~Amer-Yahia.
\newblock Discovering {C}haracterizing {R}egions for {C}onsumer {P}roducts.
\newblock In {\em Proc. DSAA}, 2015.

\bibitem{DBLP:conf/icdt/PasquierBTL99}
N.~Pasquier, Y.~Bastide, R.~Taouil, and L.~Lakhal.
\newblock Discovering {F}requent {C}losed {I}temsets for {A}ssociation {R}ules.
\newblock In {\em Proc. ICDT}, pages 398--416, 1999.

\bibitem{conf/dmkd/PeiHM00}
J.~Pei, J.~Han, and R.~Mao.
\newblock Closet: {A}n {E}fficient {A}lgorithm for {M}ining {F}requent {C}losed
  {I}temsets.
\newblock In {\em Proc. SIGMOD}, pages 21--30, 2000.

\bibitem{Piatetsky1991KDD}
G.~Piatetsky-Shapiro.
\newblock {\em Knowledge {D}iscovery in {D}atabases}.
\newblock Menlo Park, CA: AAI/MIT, 1991.

\bibitem{sokal58}
R.~R. Sokal and C.~D. Michener.
\newblock A {S}tatistical {M}ethod for {E}valuating {S}ystematic
  {R}elationships.
\newblock {\em Univ. Kans. Sci. Bull.}, 38:1409--1438, 1958.

\bibitem{datamining}
P.-N. Tan, M.~Steinbach, and V.~Kumar.
\newblock {\em Introduction to Data Mining, (First Edition)}.
\newblock W. W. Norton \& Company, 2007.

\bibitem{hbase}
{The Apache Software Foundation}.
\newblock {HB}ase.
\newblock \url{http://hbase.apache.org}.
\newblock [Online; accessed 29-Feb-2016].

\bibitem{DBLP:conf/fimi/UnoKA04}
T.~Uno, M.~Kiyomi, and H.~Arimura.
\newblock {LCM} ver. 2: {E}fficient {M}ining {A}lgorithms for
  {F}requent/{C}losed/{M}aximal {I}temsets.
\newblock In {\em Proc. ICDM Workshop FIMI}, 2004.

\bibitem{yarn}
V.~K. Vavilapalli, A.~C. Murthy, C.~Douglas, S.~Agarwal, M.~Konar, R.~Evans,
  T.~Graves, J.~Lowe, H.~Shah, S.~Seth, B.~Saha, C.~Curino, O.~O'Malley,
  S.~Radia, B.~Reed, and E.~Baldeschwieler.
\newblock Apache {H}adoop {YARN}: {Y}et {A}nother {R}esource {N}egotiator.
\newblock In {\em Proc. SOCC}, pages 5:1--5:16, 2013.

\bibitem{spark}
M.~Zaharia, M.~Chowdhury, M.~J. Franklin, S.~Shenker, and I.~Stoica.
\newblock Spark: {C}luster {C}omputing with {W}orking {S}ets.
\newblock In {\em Proc. {HotCloud}}, pages 10--10, 2010.

\end{thebibliography}
\end{document}